\pdfoutput=1
\newcommand*{\ATLASLATEXPATH}{latex/}

\documentclass[cernpreprint,texlive=2016,txfonts,UKenglish]{latex/atlasdoc} 

\usepackage{\ATLASLATEXPATH atlasphysics}
\usepackage{\ATLASLATEXPATH atlascontribute}
\usepackage{\ATLASLATEXPATH atlaspackage}
\usepackage{\ATLASLATEXPATH atlascover}
\usepackage{\ATLASLATEXPATH atlasbiblatex}
\newcommand{\hdamp}{\ensuremath{h_{\mathrm damp}}}

\newcommand{\HDampMT}{\ensuremath{h_{\mathrm damp}\!=\!m_{\mathrm{top}}}}

\usepackage{mathrsfs}
\usepackage{tabularx}
\usepackage{multirow}
\usepackage{multicol}
\usepackage{amsmath}
\usepackage{url}
\usepackage{authblk}
\usepackage{placeins}
\usepackage{hyperref}
\usepackage{\ATLASLATEXPATH adjcalc}
\usepackage{\ATLASLATEXPATH adjustbox}
\usepackage{\ATLASLATEXPATH stackengine}
\usepackage{\ATLASLATEXPATH collectbox}

\newcommand\brabar{\scalebox{.3}{(}\raisebox{-1.7pt}{\textasciitilde{}}\scalebox{.3}{)}} 

\graphicspath{{logos/}{fig/}}

\hypersetup{pdftitle={Measurements of charge and CP asymmetries in $b$-hadron decays using top-quark events collected by the ATLAS detector in $pp$ collisions at $\sqrt{s}=8~\tev{}$},pdfauthor={The ATLAS Collaboration}}

\AtlasTitle{Measurements of charge and CP asymmetries in $b$-hadron decays using top-quark events collected by the ATLAS detector in $pp$ collisions at $\sqrt{s}=8~\tev{}$}

\AtlasAbstract{
Same- and opposite-sign charge asymmetries are measured in lepton+jets~\ttbar{} events in which a $b$-hadron decays semileptonically to a soft muon, using data corresponding to an integrated luminosity of 20.3~\ifb{} from proton--proton collisions at a centre-of-mass energy of $\sqrt{s}=8~\tev{}$ collected with the ATLAS detector at the Large Hadron Collider at CERN. The charge asymmetries are based on the charge of the lepton from the top-quark decay and the charge of the soft muon from the semileptonic decay of a $b$-hadron and are measured in a fiducial region corresponding to the experimental acceptance. Four CP asymmetries (one mixing and three direct) are measured and are found to be compatible with zero and consistent with the Standard Model. 
}

\PreprintIdNumber{CERN-EP-2016-221}
\author{The ATLAS Collaboration}
\AtlasJournal{JHEP}
\AtlasJournalRef{JHEP 02 (2017) 071}
\AtlasDOI{10.1007/JHEP02(2017)071}

\begin{document}
\sloppy
\maketitle
\clearpage
\tableofcontents
\clearpage

\section{Introduction}
\label{sec:Introduction}

The violation of the combined charge conjugation (C) and parity transformation (P) of particles and antiparticles implies that the laws of physics are not the same for matter and antimatter. The study of CP violation, first observed in neutral kaon decays in 1964 by Christenson, Cronin, Fitch and Turlay~\cite{Christenson:1964fg}, has a long and rich history within particle physics.  Since the first observation, CP violation in mixing has also been observed in neutral $B_d$ and $B_s$ mesons~\cite{Aubert:2001nu,Abe:2001xe,Aaij:2013iua}, and direct CP violation has been observed in kaon decays~\cite{Burkhardt:1988yh,Fanti:1999nm,AlaviHarati:1999xp} and $B$-meson decays~\cite{Aubert:2004qm,Chao:2004mn,Poluektov:2010wz,delAmoSanchez:2010ji,Aaij:2012kz}.  However, these observations are not sufficient to explain the matter--antimatter asymmetry in the universe~\cite{Sakharov:1967dj}. In addition, a sizeable inclusive like-sign dimuon charge asymmetry (using a sample of primarily $b\overline{b}$ pairs) has been reported~\cite{Abazov:2013uma} by the D0 experiment in which an excess was observed over that predicted by the Standard Model (SM), when the measurement is interpreted in the form of CP asymmetries relevant to this paper.  This result is not confirmed by recent measurements of CP violation in $B^{0}_{d,s}$ mixing from LHCb~\cite{Aaij:2014nxa,Aaij:2016yze} and BaBar~\cite{Lees:2014kep}, which are consistent with the SM.
 
The abundance of top quarks produced at the Large Hadron Collider (LHC)~\cite{Evans:2008zzb} makes it possible to perform measurements of CP asymmetries in heavy-flavour mixing and decay from top-quark decay products, as outlined in Ref.~\cite{Gedalia:2012sx}. All existing analyses of CP violation in $B$-meson decays rely on a resonant production of $b\overline{b}$ pairs or hadroproduction. The unique aspect presented in this paper is the method by which the charge of the $b$-quark is determined, both at production and at decay. The top quark decays before hadronisation, predominantly via $t\rightarrow Wb$. In the case where the $W$-boson decays leptonically, the charge of the lepton determines the charge of the produced $b$-quark. The $b$-quark hadronises and in the case that the resulting $b$-hadron decays semileptonically, the charge of the soft lepton determines the $b$-quark charge at decay. This paper exploits a soft-muon heavy-flavour tagging (SMT) algorithm~\cite{Aad:2015ydr,ATLAS-CONF-2012-131} in order to  measure muons that originate from such a process, hereafter referred to as SMT muons. In principle this process may also be observed via soft electrons or soft $\tau$ leptons; however, in practice the reconstruction efficiency and background rejection for these channels is unfavourable when compared to soft muons.

This measurement uses data corresponding to an integrated luminosity of 20.3~\ifb{}~\cite{CERN-EP-2016-117} from proton--proton collisions at a centre-of-mass energy of $\sqrt{s}=8~\tev{}$ collected with the ATLAS~\cite{Aad:2008zzm} detector at the Large Hadron Collider at CERN. It selects events with exactly one lepton from a $W$-boson decay and at least four jets, one of which must be tagged with both a displaced-vertex $b$-tagging algorithm and the SMT algorithm. Experimental ambiguity in determining the charge of the produced $b$-quark arises in~\ttbar{} pair production events when establishing if an SMT muon originated from the same- or different-top quark as the leptonically decaying $W$-boson. This can be resolved experimentally and corrected for as part of the unfolding procedure. An illustration of same- and different-top SMT muons is shown in Figure~\ref{fig:AnalysisOutline:SameDiffTop}. In the case of a same-top SMT muon, a positively (negatively) charged $W$-boson lepton (the lepton from the decay of the $W$-boson) implies that the charge of the produced $b$-quark was negative (positive). In the case of a different-top SMT muon, a positively (negatively) charged $W$-boson lepton implies that the charge of the produced $b$-quark was also positive (negative).

\begin{figure}[!htb]
  \subfloat[Same-top SMT muon\label{fig:AnalysisOutline:SameTop}]{
    \includegraphics[width=0.45\textwidth]{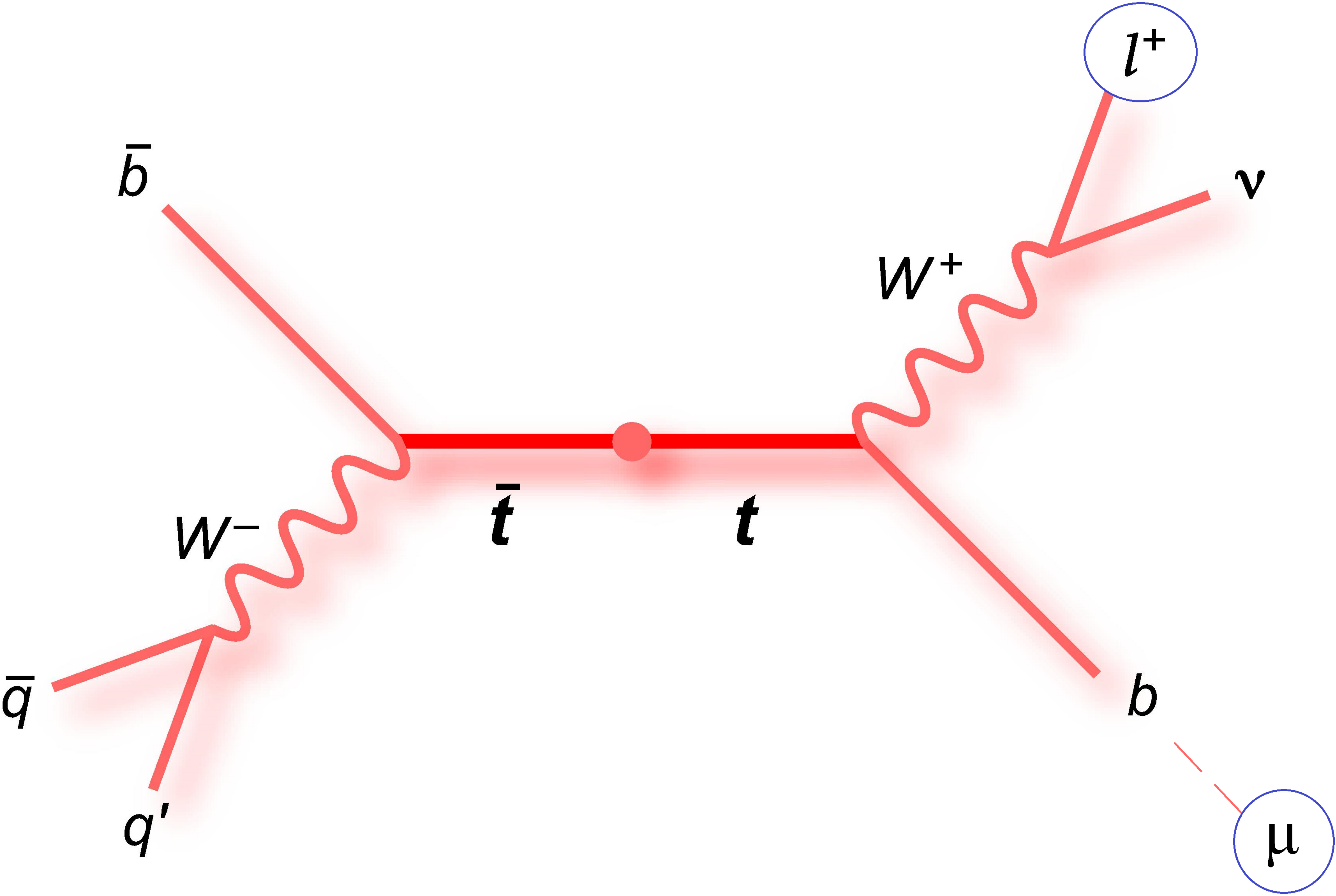}
  }
  \hfill
  \subfloat[Different-top SMT muon\label{fig:AnalysisOutline:DiffTop}]{
    \includegraphics[width=0.45\textwidth]{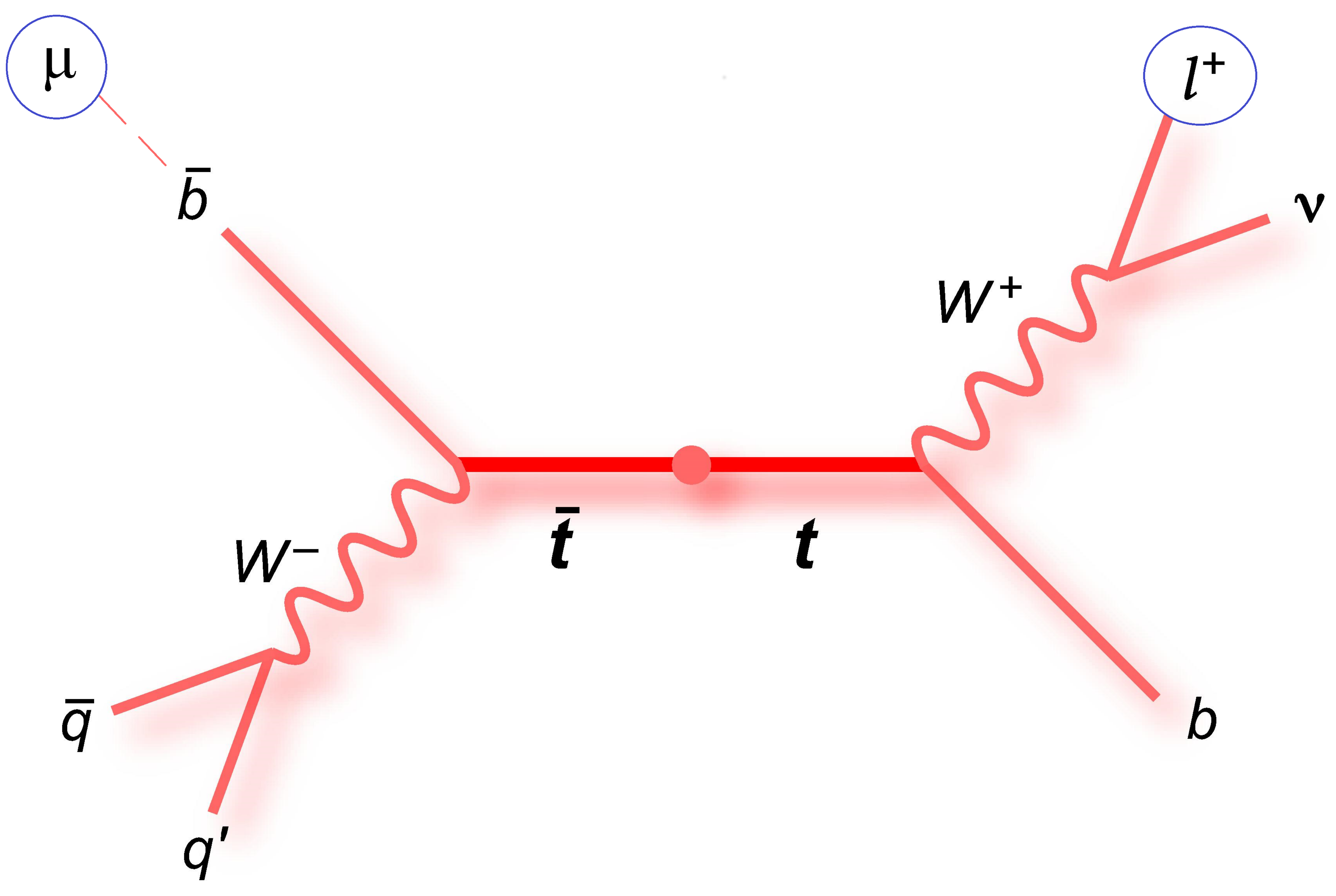}
  }
  \caption{Illustration of same- and different-top SMT muons.}
  \label{fig:AnalysisOutline:SameDiffTop}
\end{figure}

There are, assuming charge conjugation, three classes of ~\ttbar{} decay chains that produce two leptons of the same sign, given by Equations~(\ref{eg:Introduction:TopTo2Lep_same1})--(\ref{eg:Introduction:TopTo2Lep_same3}), and three classes of decay chains which produce two leptons of opposite sign, given by Equations~(\ref{eg:Introduction:TopTo2Lep_diff1})--(\ref{eg:Introduction:TopTo2Lep_diff3}) (where in each case $N_{{\stackon[.1pt]{r}{\brabar}}_{i}}$ represents the number of SMT muons from the appropriate decay channel in a well-defined fiducial region):

\begin{alignat}{2}
&N_{r_{b}} && = N\left[t  \rightarrow \ell^{+}\nu\left(b\rightarrow \overline{b}\right) \rightarrow \ell^{+}\ell^{+}X \right], \label{eg:Introduction:TopTo2Lep_same1} \\
&N_{r_{c}} &&= N\left[t  \rightarrow\ell^{+}\nu\left(b\rightarrow c\right) \rightarrow \ell^{+}\ell^{+}X \right], \label{eg:Introduction:TopTo2Lep_same2} \\
&N_{r_{c\overline{c}}} &&= N\left[t \rightarrow \ell^{+}\nu\left(b\rightarrow \overline{b} \rightarrow c\overline{c}\right) \rightarrow \ell^{+}\ell^{+}X \right], \label{eg:Introduction:TopTo2Lep_same3} \\
&N_{\widetilde{r}_{b}} &&= N\left[t \rightarrow \ell^{+}\nu b \rightarrow \ell^{+}\ell^{-}X \right], \label{eg:Introduction:TopTo2Lep_diff1} \\
&N_{\widetilde{r}_{c}} &&= N\left[t \rightarrow \ell^{+}\nu\left(b\rightarrow \overline{b} \rightarrow \overline{c}\right) \rightarrow \ell^{+}\ell^{-}X \right], \label{eg:Introduction:TopTo2Lep_diff2} \\
&N_{\widetilde{r}_{c\overline{c}}}&&= N\left[t \rightarrow \ell^{+}\nu\left(b\rightarrow  c\overline{c}\right) \rightarrow \ell^{+}\ell^{-}X \right]. \label{eg:Introduction:TopTo2Lep_diff3}
\end{alignat}

Experimentally observable charge asymmetries are formed by considering the relative difference in the probability for an initial $b$- or $\overline{b}$-quark to decay via either a positively or negatively charged SMT muon. Let $N^{\alpha\beta}$ represent the number of SMT muons observed with a charge $\beta$ in conjunction with a $W$-boson lepton of charge $\alpha$, where $\alpha,\beta=\pm1$. In the case that an SMT muon is estimated to have originated from the different top-quark to the $W$-boson lepton, the sign of the $W$-boson lepton, $\alpha$, is flipped in order to consistently represent the charge of the $b$-quark at production in both scenarios. In the case of events where both $b$-hadrons decay semileptonically and are both experimentally tagged, the event contributes twice to the asymmetries. A total of four different probabilities are considered:

\begin{alignat}{4}
& P\left(b \rightarrow \ell^{+} \right) &&=\dfrac{N\left(b \rightarrow \ell^{+}\right)}{N\left(b \rightarrow \ell^{-}\right) + N\left(b \rightarrow \ell^{+}\right)}              && = \dfrac{ N^{++} } {N^{+-} + N^{++}}  && = \dfrac{N^{++}}{N^{+}}, \\[0.5ex]
& P\left(\overline{b} \rightarrow \ell^{-}\right) &&= \dfrac{N\left(\overline{b} \rightarrow \ell^{-}\right)}{N\left(\overline{b} \rightarrow \ell^{-}\right) + N\left(\overline{b} \rightarrow \ell^{+}\right)}     && = \dfrac{ N^{--} } {N^{--} + N^{-+}} && = \dfrac{N^{--}}{N^{-}},\\[0.5ex]
& P\left(b \rightarrow \ell^{-}\right) && = \dfrac{N\left(b \rightarrow \ell^{-}\right)}{N\left(b \rightarrow \ell^{-}\right) + N\left(b \rightarrow \ell^{+}\right)}                           && = \dfrac{ N^{+-}} {N^{+-} + N^{++}}  && = \dfrac{N^{+-}}{N^{+}},\\[0.5ex]
& P\left(\overline{b} \rightarrow \ell^{+}\right) &&= \dfrac{N\left(\overline{b} \rightarrow \ell^{+} \right)}{N\left(\overline{b} \rightarrow \ell^{-}\right) + N\left(\overline{b} \rightarrow \ell^{+}\right)} && =\dfrac{ N^{-+} }{N^{--} + N^{-+}} && = \dfrac{N^{-+}}{N^{-}},
\end{alignat}

where $N^{+}\equiv N^{++}+N^{+-}$ and $N^{-}\equiv N^{-+}+N^{--}$ represent the total number of positively and negatively charged $W$-boson leptons respectively. Observable same- and opposite-sign charge asymmetries may be formed from the probabilities:

\begin{alignat}{4}
&A^{\mathrm{ss}} &&= \dfrac{P\left(b \rightarrow \ell^{+} \right) - P\left(\overline{b} \rightarrow \ell^{-}\right)}{P\left(b \rightarrow \ell^{+} \right) + P\left(\overline{b} \rightarrow \ell^{-}\right)}, \qquad &&A^{\mathrm{os}} &&= \dfrac{P\left(b \rightarrow \ell^{-}\right) - P\left(\overline{b} \rightarrow \ell^{+}\right)}{P\left(b \rightarrow \ell^{-}\right) + P\left(\overline{b} \rightarrow \ell^{+}\right)}, \label{Eq:Ass}\\[0.5ex]
&A^{\mathrm{ss}} &&= \dfrac{\left( \dfrac{N^{++}}{N^{+}} - \dfrac{N^{--}}{N^{-}} \right)}{\left( \dfrac{N^{++}}{N^{+}} + \dfrac{N^{--}}{N^{-}} \right)}, && A^{\mathrm{os}} &&= \dfrac{\left( \dfrac{N^{+-}}{N^{+}} - \dfrac{N^{-+}}{N^{-}} \right)}{\left( \dfrac{N^{+-}}{N^{+}} + \dfrac{N^{-+}}{N^{-}} \right)}. \label{Eq:Aos}
\end{alignat}

The charge asymmetries, $A^{\mathrm{ss}}$ and $A^{\mathrm{os}}$, are expressed as ratios of probabilities as this ensures that the measurements are independent of any asymmetry that could lead to a different rate of positively or negatively charged $W$-boson leptons being reconstructed. These effects can come about due to~\ttbar{} pair production asymmetries, reconstruction asymmetries or background asymmetries.  Non-~\ttbar{} backgrounds, estimated from simulation and by data-driven techniques, are subtracted from the data. The data are then unfolded to a well-defined fiducial region from which the charge asymmetries are measured. The use of a fiducial region provides a prescription, described below, in which the CP asymmetries may be extracted from the charge asymmetries, as well as reducing the experimental uncertainties by minimising the extrapolation from the reconstruction-level selection to the particle level. A more traditional dilution approach would, in this case, be able to measure the charge asymmetries but would then be unable to extract the CP asymmetries. 

The charge asymmetries are related to the CP asymmetries [Equations~(\ref{eq:interpretationCPV:AblMix})--(\ref{eq:interpretationCPV:AbcDir})] via:

\begin{alignat}{2}
& A^{\mathrm{ss}} &&= r_{b}A^{b\ell}_{\mathrm{mix}} + r_{c}\left(A^{bc}_{\mathrm{dir}} - A^{c\ell}_{\mathrm{dir}}\right) + r_{c\overline{c}}\left(A^{bc}_{\mathrm{mix}}-A^{c\ell}_{\mathrm{dir}}\right) \label{eq:Introduction:Ass} \\[0.5ex]
& A^{\mathrm{os}} &&= \widetilde{r}_{b}A^{b\ell}_{\mathrm{dir}} + \widetilde{r}_{c}\left(A^{bc}_{\mathrm{mix}} +A^{c\ell}_{\mathrm{dir}}\right) + \widetilde{r}_{c\overline{c}}A^{c\ell}_{\mathrm{dir}} \label{eq:Introduction:Aos}
\end{alignat}

where the decay-chain fractions, $r_{i}$ and $\widetilde{r}_{i}$, represent the relative rates of each channel. The decay-chain fractions are dependent on the fiducial region chosen and are calculated as:

\begin{alignat}{5}
& r_{b} &&= \dfrac{ N_{r_{b}} } { N_{r_{b}} + N_{r_{c}} + N_{r_{c\overline{c}}} }, && \qquad && \widetilde{r}_{b} &&= \frac{ N_{\widetilde{r}_{b}} } { N_{\widetilde{r}_{b}} + N_{\widetilde{r}_{c}} + N_{\widetilde{r}_{c\overline{c}}} },\\[0.5ex]
& r_{c} &&= \frac{ N_{r_{c}} } { N_{r_{b}} + N_{r_{c}} + N_{r_{c\overline{c}}} }, && \qquad && \widetilde{r}_{c} &&= \frac{ N_{\widetilde{r}_{c}} } {  N_{\widetilde{r}_{b}} + N_{\widetilde{r}_{c}} + N_{\widetilde{r}_{c\overline{c}}} }, \\[0.5ex]
& r_{c\overline{c}} &&= \frac{ N_{r_{c\overline{c}}} } { N_{r_{b}} + N_{r_{c}} + N_{r_{c\overline{c}}} }, && \qquad && \widetilde{r}_{c\overline{c}} &&= \frac{ N_{\widetilde{r}_{c\overline{c}}}}{ N_{\widetilde{r}_{b}} + N_{\widetilde{r}_{c}} + N_{\widetilde{r}_{c\overline{c}}} }.
\end{alignat}

The CP asymmetries related to $B_{q}-\overline{B}_{q}$ mixing and direct CP-violating $b$- and $c$-decays are defined as:

\begin{alignat}{3}
&A^{b\ell}_{\mathrm{mix}} = \dfrac{ \Gamma \left(b \rightarrow \overline{b} \rightarrow \ell^{+} X \right) - \Gamma \left(\overline{b} \rightarrow b \rightarrow \ell^{-} X \right)}{ \Gamma \left(b \rightarrow \overline{b} \rightarrow \ell^{+} X \right) + \Gamma \left(\overline{b} \rightarrow b \rightarrow \ell^{-} X \right)}, \label{eq:interpretationCPV:AblMix} \\[1ex]
&A^{bc}_{\mathrm{mix}} = \dfrac{ \Gamma \left(b \rightarrow \overline{b} \rightarrow \overline{c} X \right) - \Gamma \left(\overline{b} \rightarrow b \rightarrow c X \right)}{ \Gamma \left(b \rightarrow \overline{b} \rightarrow \overline{c} X \right) + \Gamma \left(\overline{b} \rightarrow b \rightarrow c X \right)}, \label{eq:interpretationCPV:AbcMix} \\[1ex]
&A^{b\ell}_{\mathrm{dir}} = \dfrac{ \Gamma \left(b \rightarrow \ell^{-} X \right) - \Gamma \left(\overline{b} \rightarrow \ell^{+} X \right)}{ \Gamma \left(b \rightarrow \ell^{-} X \right) + \Gamma \left(\overline{b} \rightarrow \ell^{+} X \right)}, \label{eq:interpretationCPV:AblDir} \\[1ex]
&A^{c\ell}_{\mathrm{dir}} = \dfrac{ \Gamma \left(\overline{c} \rightarrow \ell^{-} X_{L} \right) - \Gamma \left(c \rightarrow \ell^{+} X_{L} \right)}{ \Gamma \left(\overline{c} \rightarrow \ell^{-} X_{L} \right) + \Gamma \left(c \rightarrow \ell^{+} X_{L} \right)}, \label{eq:interpretationCPV:AclDir} \\[1ex]
&A^{bc}_{\mathrm{dir}} = \dfrac{ \Gamma \left(b \rightarrow c X_{L} \right) - \Gamma \left(\overline{b} \rightarrow \overline{c} X_{L} \right)}{ \Gamma \left(b \rightarrow c X_{L} \right) + \Gamma \left(\overline{b} \rightarrow \overline{c} X_{L} \right)}, \label{eq:interpretationCPV:AbcDir}
\end{alignat}

where $X \left(X_{L}\right)$ denotes an inclusive hadronic final state with no leptons, and with both light and charm quarks (with light quarks only).

\section{The ATLAS detector}
\label{sec:Detector}

The ATLAS~\cite{Aad:2008zzm} detector at the LHC~\cite{Evans:2008zzb} covers the pseudorapidity\footnote{ATLAS uses a right-handed coordinate system with its origin at the nominal interaction point (IP) in the centre of the detector and the $z$-axis along the beam pipe. The $x$-axis points from the IP to the centre of the LHC ring, and the $y$-axis points upwards. Cylindrical coordinates $(r,\phi)$ are used in the transverse plane, $\phi$ being the azimuthal angle around the $z$-axis. The pseudorapidity is defined in terms of the polar angle $\theta$ as $\eta = -\ln \tan(\theta/2)$. Angular distance is measured in units of $\Delta R \equiv \sqrt{(\Delta\eta)^{2} + (\Delta\phi)^{2}}$.} range $|\eta|<4.9$ and the full azimuthal angle $\phi$. It consists of the following main subsystems: an inner tracking system immersed in a 2 T magnetic field provided by a superconducting solenoid, electromagnetic (EM)  and hadronic calorimeters, and a muon spectrometer incorporating three large superconducting toroid magnets composed of eight coils each. The inner detector (ID) is composed of three subsystems: the pixel detector, the semiconductor tracker and the transition radiation tracker. The ID provides tracking information in the pseudorapidity range $|\eta|<2.5$, calorimeters measure energy deposits (clusters) for $|\eta|<4.9$, and the muon spectrometer records tracks within $|\eta|<2.7$. A three-level trigger system~\cite{Aad:2012xs} is used to select interesting events. It consists of a level-1 hardware trigger, reducing the event rate to at most 75 kHz, followed by two software-based trigger levels, collectively referred to as the high-level trigger, yielding a recorded event rate of approximately 400 Hz on average, depending on the data-taking conditions.

\section{Object and event selection}
\label{sec:ObjSel}

The results are based on proton--proton collisions collected with the ATLAS experiment at the LHC at a centre-of-mass energy of $\sqrt{s}=8~\tev{}$ in 2012. The total integrated luminosity available for the analysis is 20.3~\ifb{}.
Only events collected under stable beam conditions with all relevant detector subsystems operational are used. Events are selected using single-lepton triggers with transverse momentum $\left(\pt{}\right)$ thresholds of 24 or 60~\gev{} for electrons and 24 or 36~\gev{} for muons. The triggers with the lower \pt{} threshold include isolation requirements on the candidate lepton in order to reduce the trigger rate to an acceptable level. To ensure that events originate from proton--proton collisions, events are required to have at least one reconstructed vertex with at least five associated tracks of $\pt{}>400~\mev{}$. If more than one vertex is found, the hard-scatter primary vertex is identified as the one which has the largest sum of the squared transverse momenta of its associated tracks.

A brief summary of the main reconstruction and identification criteria applied for each physics object is given below.

\subsection{Reconstruction-level objects and event selection}
Electron candidates~\cite{Aad:2014fxa} from $W$-boson decay are reconstructed from energy deposits in the electromagnetic calorimeter that are matched to reconstructed tracks in the inner detector. The electrons are required to have a transverse energy, $E_{\mathrm{T}}>25~\gev{}$ and $|\eta_{\mathrm{cluster}}|<2.47$, where $\eta_{\mathrm{cluster}}$ is the pseudorapidity of the electromagnetic energy cluster in the calorimeter with respect to the geometric centre of the detector. Candidates in the barrel/endcap transition region $(1.37<|\eta_{\mathrm{cluster}}|<1.52)$ are excluded. The longitudinal impact parameter of the track with respect to the primary vertex, $|z_{0}|$, is required to be less than 2 mm. Electrons must satisfy ``tight''~\cite{BenharNoccioli:2138479} quality requirements based on the shape of the energy deposit and on the match to the track, in order to distinguish them from hadrons. Additionally, isolation requirements are imposed based on nearby tracks or calorimeter energy deposits. These requirements depend on the electron kinematics and are derived to give an efficiency that is constant with respect to the electron $E_{\mathrm{T}}$ and $\eta$. The cell-based isolation uses the sum of all calorimeter cell energies within a cone of size $\Delta R = 0.2$ around the electron direction while the track-based isolation sums all track momenta within a cone of size $\Delta R = 0.3$; in both cases the contribution from the electron itself is excluded and the isolation criteria are optimised to individually result in a 90\% efficiency for electrons from $Z\rightarrow \ee$ decays.  To prevent double-counting of electron energy deposits as jets, jets within $\Delta R = 0.2$ of a reconstructed electron are removed. Finally, if the nearest jet surviving the above cut is within $\Delta R = 0.4$  of the electron, the electron is discarded to ensure it is cleanly separated from nearby jet activity.

Muon candidates~\cite{Aad:2014zya,Aad:2014rra} from $W$-boson decay are reconstructed by matching tracks formed in the muon spectrometer and the inner detector. The final candidates are refitted using the complete track information from both detector systems, and are required to have $\pt{}>25~\gev{}$, $|\eta|<2.5$, and $|z_{0}|<2$ mm. Muons are required to satisfy a ~\pt{}-dependent track-based isolation: the scalar sum of the track~\pt{} within a cone of variable size around the muon, $\Delta R = 10~\gev{}/\pt{}^{\mu}$ (excluding the muon track itself) must be less than 5\% of the muon~\pt{}, corresponding to a 97\% selection efficiency for muons from $Z \rightarrow \mu^{+}\mu^{-}$ decays.  To reduce background from heavy-flavour decays inside jets, muons are required to be separated by $\Delta R > 0.4$ from the nearest jet.

Jets are reconstructed with the anti-$k_{t}$ algorithm~\cite{Cacciari:2008gp} with a radius parameter $R = 0.4$. Locally calibrated topological clusters of calorimeter cells~\cite{Lampl:1099735,Aad:2011he} are calibrated to the energy scale of particle-level hadrons and are used as input to the jet clustering algorithm. Jets are required to have $\pt{}>25~\gev{}$ and $|\eta|<2.5$. In order to suppress jets originating from pile-up, a requirement on the jet vertex fraction (JVF)~\cite{Aad:2015ina} is applied. The JVF is defined as the summed scalar~\pt{} of tracks associated with both the reconstructed primary vertex and the jet, divided by the summed scalar~\pt{} of all tracks associated with the jet. For jets with $\pt{}<50~\gev{}$ and $|\eta|<2.4$, a JVF $>0.5$ is required.

Jets containing $b$-hadrons are identified ($b$-tagged) using two methods. Firstly, jets are required to be tagged using a displaced-vertex algorithm that uses multivariate techniques to combine information from the impact parameters of associated tracks and topological properties of secondary and tertiary decay vertices reconstructed within the jet~\cite{ATLAS-CONF-2014-046,ATLAS-CONF-2014-004,Aad:2015ydr}. The algorithm's operating point used for this measurement corresponds to a 85\% efficiency to tag $b$-quark jets and a rejection factor of 10 for light-quark- or gluon-initiated jets, as determined for jets with $\pt>20~\gev{}$ and $|\eta|<2.5$ in simulated~\ttbar{} events.  Secondly, jets are required to be tagged by the SMT algorithm~\cite{Aad:2015ydr,ATLAS-CONF-2012-131}.

The SMT algorithm identifies jets containing $b$- and $c$-quarks via the presence of one or more muons in close proximity to the jet. The momentum imbalance (MI) is defined as $\mathrm{MI}=\left(p_{\mathrm{ID}}-p_{\mathrm{ME}}\right)/p_{\mathrm{ID}}$, where $p_{\mathrm{ID}}$ is the momentum of the muon as measured by the inner detector, while $p_{\mathrm{ME}}$ is the momentum of the muon as measured by the muon spectrometer with the track extrapolated back to the primary vertex. Muons originating from heavy-flavour decays close to the primary vertex have a very small MI whilst muons from light-flavour decays of particles with long lifetimes, and therefore originating far from the primary vertex, have larger values of MI. SMT muons are selected with $\pt{}>4~\gev{}$, $|\eta|<2.5$ and an MI less than $0.1$ within a cone of size $\Delta R = 0.5$ around a selected jet.

Events are required to have exactly one candidate electron or muon from a $W$-boson decay and at least four jets. The selected $W$-boson lepton is required to match the lepton reconstructed by the high-level trigger within a cone of size $\Delta R = 0.15$. Events are required to have at least one $b$-tagged jet that was tagged by both $b$-tagging algorithms. After selecting $b$-tagged jets, the remaining jets are added in order of highest to lowest $\pt$; in the rare case of three $b$-tagged jets, the lowest-$\pt$ of the three is treated as a light jet.  Additionally, for the $\mu$+jets channel only, the invariant mass of the $W$-boson muon and the SMT muon is calculated and events with $8~\gev{}<m_{\mu\mu}<11~\gev{}$ or $80~\gev{}<m_{\mu\mu}<100~\gev{}$ are vetoed from the analysis in order to exclude events containing $\Upsilon \rightarrow \mu \mu$  and $Z\rightarrow \mu \mu$ decays, respectively.

\subsection{Particle-level objects and simulated event selection}
\label{sec:fidSel:partDef}
The particle-level definition of simulated objects is based on particles with a proper lifetime $\tau_{\mathrm{particle}}>3\times 10^{-11}\mathrm{s}$. The definitions used here follow very closely previous ATLAS~\ttbar{} fiducial definitions~\cite{Aad:2014iaa}. Fiducial requirements are placed only on jets and charged leptons.

Prompt electrons and muons, i.e. those that are not hadron decay products, are considered for the fiducial lepton definition. The four-momenta of any photon within a cone of size $\Delta R = 0.1$ around a prompt lepton is added to the four-momenta of the prompt lepton, which are hereafter referred to as dressed leptons. Dressed leptons are required to have $\pt>25~\gev{}$ and $|\eta|<2.5$.

Jets are obtained by clustering all stable particles, except the leptons, dressed with their associated photons, and neutrinos that are not hadron decay products, using the anti-$k_{t}$ algorithm with a radius parameter $R = 0.4$. Particles from the underlying event are included in this definition, whereas particles from additional inelastic proton--proton collisions (pile-up) are not included. Jets are required to have $\pt>25~\gev{}$ and $|\eta|<2.5$. 

A jet is defined as a $b$-jet by its association with one or more $b$-hadrons with $\pt>5~\gev{}$ within a cone of size $\Delta R = 0.5$ around the direction of the jet. The jet is further required to be within a cone of size $\Delta R = 0.5$ around a muon with a $\pt>4~\gev{}$ where that muon has an ancestral link to the associated $b$-hadron. The muon is further required to have an ancestral link to a top quark.

Selected events must  contain a single prompt lepton and at least four jets, of which at least one jet must be identified as a $b$-jet as defined above. In order to ensure all particle-level objects are well separated, events are rejected if any of the jets satisfying the fiducial requirements lie within a cone of size $\Delta R = 0.4$ around a prompt dressed lepton.

\section{Simulation samples and background estimates} 
\label{sec:SimSamples}

\subsection{Signal and background modelling}
The sample composition is dominated by \ttbar events. Contributions from other processes arise from $W$+jets, $Z$+jets, single top ($t$-channel, $Wt$ and $s$-channel) and diboson ($WW$, $WZ$, $ZZ$) production as well as events with one or more non-prompt or misidentified leptons from decays of hadrons. All background processes are modelled using Monte Carlo (MC) simulations except for the background from non-prompt or fake leptons, and the $W$+jets normalisation and flavour composition, which use data-driven techniques.

The nominal sample used to model~\ttbar{} events is generated using the {\normalfont \scshape{PowhegBox}} (version 1, r2330) NLO generator~\cite{Nason:2004rx,Frixione:2007vw,Alioli:2010xd,Frixione:2007nw}, with the NLO CT10 parton distribution function (PDF)~\cite{Lai:2010vv,Gao:2013xoa} assuming a top-quark mass of 172.5~\gev{}. It is interfaced to {\normalfont \scshape{Pythia}} 6.427~\cite{Sjostrand:2006za} with the CTEQ6L1~\cite{Pumplin:2002vw,Nadolsky:2008zw} PDF and the Perugia2011C~\cite{Skands:2010ak} settings for the tunable parameters (hereafter referred to as tune). The \hdamp\,parameter of {\normalfont \scshape{PowhegBox}}, which controls the~\pt{} of the first additional emission beyond the Born configuration, is set to $m_{\mathrm{top}}=172.5~\gev{}$. The main effect of this is to regulate the high-\pt{} emission against which the \ttbar system recoils. The \ttbar sample is normalised to the theoretical prediction of $253^{+13}_{-15}~\mathrm{pb}$ calculated at next-to-next-leading order (NNLO) in QCD that includes resummation of next-to-next-leading logarithmic (NNLL) soft gluon terms with {\normalfont \scshape{Top++2.0}}~\cite{Beneke:2011mq,Cacciari:2011hy,Baernreuther:2012ws,Czakon:2012zr,Czakon:2012pz,Czakon:2013goa,Czakon:2011xx}. The quoted uncertainty includes the scale uncertainty and the uncertainties from PDF and~\alphas{} choices. 

Background samples of single top quark events corresponding to the $t$-channel, $s$-channel and $Wt$ production mechanisms are generated with {\normalfont \scshape{PowhegBox}} (version 1, r2330)~\cite{Alioli:2009je,Re:2010bp} using the CT10 PDF set. All samples are interfaced to {\normalfont \scshape{Pythia}} 6.427 with the CTEQ6L1 set of parton distribution functions and the Perugia2011C tune. Overlaps between the \ttbar and $Wt$ final states are removed according to the inclusive Diagram Removal scheme~\cite{Frixione:2008yi}. The single-top quark samples are normalised to the approximate NNLO theoretical cross-sections~\cite{Kidonakis:2011wy,Kidonakis:2010tc,Kidonakis:2010ux} using the MSTW2008 NNLO PDF set~\cite{Martin:2009iq}.

Samples of $WW$/$WZ$/$ZZ$+jets are generated using the {\normalfont \scshape{Alpgen}} v2.14~\cite{Mangano:2002ea} leading-order (LO) generator and the MSTW2008 NLO PDF set for all decay channels. Parton shower and fragmentation are modelled with {\normalfont \scshape{Herwig}} 6.520~\cite{Corcella:2000bw}. All diboson samples are normalised to their NLO theoretical cross-sections~\cite{Campbell:1999ah,Campbell:2011bn}, as calculated with MCFM~\cite{Campbell:2003hd}.

Samples of $W$+jets and $Z/\gamma^{*}$+jets are generated using {\normalfont \scshape{Alpgen}} v2.14 and the CTEQ6L1 PDF set. Parton shower and fragmentation are modelled with {\normalfont \scshape{Pythia}} 6.426. To avoid double-counting of partonic configurations generated by both the matrix-element calculation and the parton-shower evolution, a parton--jet matching scheme (``MLM matching'')~\cite{Mangano:2001xp} is employed. The $W/Z$+jets samples are generated with up to five additional partons, separately for production in association with $b$-quarks, $c$-quarks and light quarks. The overlap between events with heavy-flavour quarks obtained from the matrix element and the parton shower is removed. The $W/Z$+jets backgrounds are normalised to the inclusive NNLO theoretical cross-sections~\cite{Melnikov:2006kv}.

The samples that use {\normalfont \scshape{Pythia}} or {\normalfont \scshape{Herwig}}~\cite{Frixione:2002ik,Frixione:2003ei} for showering and hadronisation are interfaced to {\normalfont \scshape{Photos}}~\cite{Golonka:2005pn} for modelling of the QED final-state radiation and {\normalfont \scshape{Tauola}}~\cite{Jadach:1990mz} for modelling of the decays of $\tau$ leptons. All samples are simulated taking into account the effects of multiple proton--proton interactions based on the pile-up conditions in the 2012 data. The pile-up interactions are modelled by overlaying simulated hits from events with exactly one inelastic collision per bunch crossing with hits from minimum-bias events that are produced with {\normalfont \scshape{Pythia}} 8.160~\cite{Sjostrand:2007gs} using the A2M tune~\cite{ATL-PHYS-PUB-2012-003} and the MSTW2008 LO PDF set. All event generators using {\normalfont \scshape{Herwig}} are also interfaced to {\normalfont \scshape{Jimmy}} v4.31~\cite{Butterworth:1996zw} to simulate the underlying event. 

All simulated samples used for the measurement are processed through a simulation~\cite{Aad:2010ah} of the detector geometry and response using {\normalfont \scshape{Geant4}}~\cite{Agostinelli:2002hh}. All simulated samples are processed through the same reconstruction software as the data. Simulated events are corrected so that the object identification efficiencies, energy scales and energy resolutions match those determined in data control samples. The alternative \ttbar samples used for modelling uncertainties, described in Section~\ref{sec:syst:model}, are instead processed with the ATLFAST-II~\cite{Aad:2010ah} simulation. This employs a parameterisation of the response of the electromagnetic and hadronic calorimeters, and {\normalfont \scshape{Geant4}} for the other detector components. The $b$-hadrons are decayed in either {\normalfont \scshape{Pythia}} or {\normalfont \scshape{Herwig}}, and all samples are reweighted, as detailed in Section~\ref{sec:syst:model}, such that the $b$-hadron production and the hadron-to-muon branching ratios match the values found in the Particle Data Group's review of particle physics (RPP)~\cite{Agashe:2014kda}.

\subsection{$W$+jets normalisation}
The $W$+jets process at the LHC is charge asymmetric, with the ratio of cross-sections much better predicted theoretically than either of the individual cross-sections~\cite{Martin:2009iq}. The overall normalisation and the additional jets' flavour composition, $\left(bb,cc,c,\mathrm{light\,flavours}\right)$, of the $W$+jets background are determined via a data-driven method~\cite{Aad:2012qf} using an orthogonal 1-lepton + 2-jets exclusive region, which is dominated by the $W$+jets process. The charge asymmetry normalisation and flavour composition are then extrapolated into the 1-lepton + $\geq$ 4-jets inclusive region used in this measurement. This estimation is performed in such a way that each variation reflecting a considered systematic uncertainty uses its own associated $W$+jets estimate, with only the statistical component of the estimate considered separately.

\subsection{Backgrounds with fake or non-prompt leptons}
Events with no prompt leptons may satisfy the selection criteria if one or more jets are misidentified as isolated leptons, or if the jets include hadrons decaying to leptons which then satisfy lepton identification and isolation requirements. Such cases are hereafter referred to as fake leptons.

This background is estimated from data using a matrix method~\cite{ATLAS-CONF-2014-058}. A sample enhanced in fake leptons is selected by removing the lepton isolation requirements and, for electrons, loosening the identification criteria (these requirements are detailed in Section~\ref{sec:ObjSel}). Next, the efficiency for these ``loose'' leptons to satisfy the tight criteria is measured in data, separately for prompt and fake leptons. For prompt leptons it is taken from a sample of $Z$-boson decays, while for fake leptons it is estimated from events with low missing transverse momentum or high lepton impact parameter. This information, taken together, allows the number of fake leptons satisfying the tight criteria to be calculated.

\section{Systematic uncertainties} 
\label{sec:Uncertainties}
Several sources of systematic uncertainty are considered that can affect the normalisation of signal and background and/or the shape of the relevant distributions. Individual sources of systematic uncertainty are considered to be uncorrelated. Correlations of a given systematic uncertainty are maintained across signal and background processes and channels. Sources of systematic uncertainty are separated into experimental terms, which do not affect the decay-chain fractions, and into modelling terms which have a direct impact on the decay-chain fractions.

\subsection{Experimental uncertainties}
\label{sec:Uncertainties:expt}

The experimental uncertainties considered are those that affect the reconstruction and selecton of leptons and jets. For the measurements of the charge and CP asymmetries, some additional effects are also taken into account. Those pertain to the SMT muon identification and associated mistag, as described below. Due to the charge asymmetries being defined as a ratio of probabilities, as described in Section~\ref{sec:Introduction}, the analysis is not sensitive to experimental effects which could differently affect the positively and negatively charged $W$-boson leptons. 

\textbf{Lepton reconstruction, identification and trigger:} The charge-dependent reconstruction and identification efficiency of electrons and muons, their isolation, as well as the efficiency of the triggers used to record the event, differ slightly between data and simulation. Tag-and-probe techniques, using $Z \rightarrow \ee$, $Z \rightarrow \mu^{+}\mu^{-}$ and $\jpsi\rightarrow \mu^{+}\mu^{-}$ data and simulation~\cite{Aad:2014fxa,Aad:2014rra,BenharNoccioli:2138479}, are used to correct the simulation efficiencies, and the uncertainties associated with the tag-and-probe techniques are propagated through to the results. 

\textbf{Lepton momentum scale and resolution:} The lepton momentum scale and resolution differ slightly between data and simulation and are corrected for, according to the charge of the lepton, by using reconstructed distributions of the $Z \rightarrow \ell^{+}\ell^{-}$ and $\Jpsi \rightarrow \ell^{+}\ell^{-}$ masses~\cite{Aad:2014nim,Aad:2014rra}. In the case of electrons, $E/p$ studies using $W \rightarrow e\nu$ events are also used. In the case of muons, momentum scale and resolution corrections are only applied to the simulation, while for electrons these corrections are applied to both data and simulation. Uncertainties in both the momentum scale and resolution in the muon spectrometer and the tracking systems are considered, and varied separately. 

\textbf{Lepton charge misidentification:} Charge misidentification of an electron occurs if an isolated prompt electron is reconstructed with a wrong charge assignment. The misidentification is mostly caused by the emission of bremsstrahlung under a small angle with a subsequent conversion of the emitted photon and the misassociation of one of the conversion tracks with the cluster of the original electron. In addition, for high $E_{\mathrm{T}}$ and therefore increasingly straight tracks, charge misidentification can be caused by a failure to correctly determine the curvature of the track associated to the electron. The charge misidentification is estimated in data and simulation using $Z\rightarrow\ee$ tag-and-probe methods~\cite{BenharNoccioli:2138479}. The charge misidentification probability is correlated with the amount of bremsstrahlung and thus with the amount of traversed material; the probabilities are quite low in the central region of the detector but can reach almost 3\% for very high values of $|\eta|$. However, the fraction of  selected electrons that are potentially affected by this higher charge misidentification is very small. The possibility of muon charge misidentification is extremely small and is considered negligible. The uncertainties associated with the tag-and-probe techniques are propagated through to the results.

\textbf{Jet energy scale:} The jet energy scale (JES) and its uncertainty have been derived by combining information from test-beam data, LHC collision data and simulation~\cite{Aad:2011he,Aad:2014bia}. The jet energy scale uncertainty is split into 22 uncorrelated sources, each of which can have different~\pt{} and $\eta$ dependencies. 

\textbf{Jet energy resolution:} The jet energy resolution (JER) has been determined separately for data and simulation using two \textit{in situ} techniques~\cite{Aad:2012ag}. The fractional~\pt{} resolution for a given jet is measured as a function of its~\pt{} and $\eta$. A systematic uncertainty is defined as the difference in quadrature between the JER for data and simulation and is applied as an additional smearing to the simulation.

\textbf{Jet reconstruction efficiency:} The jet reconstruction efficiency is found to be about 0.2\% lower in the simulation than in the data for jets with~\pt{} below 30~\gev{}, and consistent with data for higher jet~\pt{}. To evaluate the systematic uncertainty due to this small difference, 0.2\% of the jets with~\pt{} below 30~\gev{} are removed randomly and all jet-related kinematic variables are recomputed. 

\textbf{Jet vertex fraction:} The efficiency for each jet to satisfy the jet vertex fraction requirement is measured in $Z\left(\rightarrow \ell^{+}\ell^{-}\right)+1$-jet events in data and simulation~\cite{Aad:2015ina}, selecting separately events enriched in hard-scatter jets and events enriched in jets from other proton interactions in the same bunch crossing (pile-up). The corresponding uncertainty is evaluated in the analysis by changing the nominal JVF requirement value.

\textbf{Flavour tagging uncertainty:} The efficiencies for $b$-jets, $c$-jets and light-flavor jets to satisfy the displaced-vertex $b$-tagging criteria have been evaluated in data, and corresponding correction factors have been derived for jets in simulation~\cite{ATLAS-CONF-2014-046,ATLAS-CONF-2014-004}. These scale factors and their uncertainties are applied to each jet depending on its flavour and~\pt{}. In the case of light-flavour jets, the corrections also depend on the jet $\eta$. Moreover, the scale factors for $\tau$-jets are set to those for $c$-jets and an additional extrapolation uncertainty is considered.  

\textbf{SMT uncertainty:} The reconstruction and identification efficiencies and uncertainties that are used for prompt muons are also used for SMT muons. In addition, tag-and-probe techniques using $Z \rightarrow \mu^{+}\mu^{-}$ and $\jpsi\rightarrow \mu^{+}\mu^{-}$ data and simulation are used to calibrate the charge-dependent efficiency of the SMT momentum imbalance selection criteria. The probability of the SMT algorithm mistagging a light-flavour jet is evaluated in data and simulation using a $W$+jets sample to obtain a sample of low-\pt{} jets which are predominantly light flavour, and correcting for the contamination from $W$+ heavy-flavour jets. An additional dijet method for high-\pt{} jets is also used, with all associated uncertainties from both methods propagated to the results. The mistag rate for positively and negatively charged muon is found to be compatible within uncertainties.

\textbf{Background uncertainties:} The data-driven fake-lepton estimate is evaluated using alternative parameterisations of the matrix method, and the uncertainties detailed in Ref.~\cite{ATLAS-CONF-2014-058} are propagated to the results. The fake lepton uncertainty on the $e$+jets channel is larger than that of the $\mu$+jets channel.  Each individual systematic uncertainty uses its own $W$+jets data-driven estimate, and a residual uncertainty based on the statistical uncertainty of the method is propagated to the results. The cross-sections of the background processes are varied within uncertainty and propagated through to the results.

\subsection{Modelling uncertainties}
\label{sec:syst:model}
\textbf{$b$-hadron production:} The production fractions for weakly decaying $b$-hadrons are documented in the RPP. All simulation samples used in this measurement are reweighted to account for differences between the RPP values and those implemented in the MC simulation. The RPP additionally provides uncertainties in the $b$-hadron production fractions as well as the correlations between the different $b$-hadrons, leading to a set of variations for these systematic uncertainties, that are propagated throughout the measurement.

\textbf{Hadron-to-muon branching ratios:} The $b$- and $c$-hadron to muon branching ratios in the simulation samples are reweighted to match those found in the RPP for the $B^{0},B^{+},B^{0}_{s}$, and $b$-baryon admixture. The uncertainties in the RPP branching ratios are propagated throughout the measurement.

\textbf{Single-top production asymmetry:} At the LHC, $t$-channel and $s$-channel single-top events are produced asymmetrically, with more top-quark events produced than antitop-quark events, whilst the $Wt$ channel single-top events are produced symmetrically. Each process, $t$-channel and $s$-channel, has a total cross-section with an uncertainty. A conservative systematic uncertainty assumes that the uncertainty in the total cross-section comes exclusively from either an excess or deficit of (anti)top-production. Correlations between top- and antitop-quark production are not considered, which would reduce the impact of these variations. A total of four variations for the two systematic uncertainties are considered for each channel, varying the (anti)top-quark production percentages up and down. This conservative systematic uncertainty has a small overall impact on the final results.

\textbf{NLO generator:} An uncertainty due to the choice of NLO generator is evaluated as the full difference, symmetrised, between ATLFAST-II~\ttbar{} samples of {\normalfont \scshape{MC@NLO}} and {\normalfont \scshape{PowhegBox}}, both interfaced with {\normalfont \scshape{Herwig}}.

\textbf{Parton shower and hadronisation:} An uncertainty due to the choice of parton shower and hadronisation models is evaluated as the full difference, symmetrised, between ATLFAST-II~\ttbar{} samples of {\normalfont \scshape{PowhegBox}} interfaced to {\normalfont \scshape{Pythia}} and to {\normalfont \scshape{Herwig}}. For both samples, the \hdamp\,parameter of {\normalfont \scshape{PowhegBox}} is set to infinity.

\textbf{Additional radiation:} The uncertainties due to the amount of initial- and final-state radiation and the factorisation and renormalisation scales, $\mu$, are evaluated together using three {\normalfont \scshape{PowhegBox}} ATLFAST-II~\ttbar{} samples. A nominal sample with $\mu=1$ and \HDampMT\,is compared to a sample with $\mu=2$ and \HDampMT\,and to a sample with $\mu=0.5$ and \hdamp$=2\times m_{\mathrm{top}}$. The uncertainty due to additional radiation is evaluated as the full difference, symmetrised, between the larger of the two comparisons.

\textbf{Parton distribution function:} The evaluation of the PDF uncertainty follows the PDF4LHC prescription~\cite{Botje:2011sn} using the CT10, MSTW2008 and NNPDF2.3~\cite{Ball:2012cx} PDF sets. This systematic uncertainty is evaluated in a full simulation~\ttbar{} sample generated with {\normalfont \scshape{MC@NLO}} using {\normalfont \scshape{Herwig}} for the parton shower, AUET2 for the underlying event tune and CT10 as the nominal PDF.

\section{Event yields and~\ttbar{} cross-section}
\label{sec:xsec}

This paper presents the measurement of charge and CP asymmetries observed in~\ttbar{} events. In order to demonstrate that the signal and relevant backgrounds are well understood, an inclusive~\ttbar{} cross-section measurement is also reported. More precise measurements of this quantity are reported elsewhere~\cite{Aad:2014kva}.  

The event yields for the $b$-tagged $e$+jets and $\mu$+jets channels, along with the expected~\ttbar{} signal and relevant backgrounds, are shown in Table~\ref{tab:EventYields:Tag}. Distributions showing the agreement between data and simulation are shown in Figure~\ref{fig:plots} for both the $e$+jets and $\mu$+jets channels, illustrating the $b$-tagged jet multiplicity, the $b$-tagged jet~\pt{} and the SMT muon~\pt{}. The predictions are found to be in good agreement with the data within uncertainties.

\begin{table}[!htb]
\begin{center}
\begin{tabular}{l|rl|rl}
                 & \multicolumn{2}{c|}{$e$+jets} & \multicolumn{2}{c}{$\mu$+jets} \\
\hline
$WW$, $WZ$ , $WW$     & 50 &$\pm$ 7            & 45 &$\pm$ 5   \\
$Z$+jets     & 800 &$\pm$ 80         & 450 &$\pm$ 60 \\
Multijet     & 1 800 &$\pm$ 1 400    & 1 500 &$\pm$ 330     \\
Single top       & 1 800 &$\pm$ 150      & 2 000 &$\pm$ 150\\
$W$+jets     & 2 500 &$\pm$ 160      & 2 800  &$\pm$ 150 \\
\ttbar{}    & 30 000 &$\pm$ 1 900        & 34 000 &$\pm$ 2 000     \\
\hline
Expected         & 37 000 &$\pm$ 2 600  & 41 000 &$\pm$ 2 300    \\
Data             & 36 796 &         & 40 807 &\\

\end{tabular}
\caption{\label{tab:EventYields:Tag}Observed and expected event yields for the $e$+jets and $\mu$+jets channels, with combined total statistical and systematic uncertainties.}
\end{center}
\end{table}

\begin{figure}[!htb]
  \vspace{-4mm}  
  \subfloat[]{
    \includegraphics[width=0.45\textwidth]{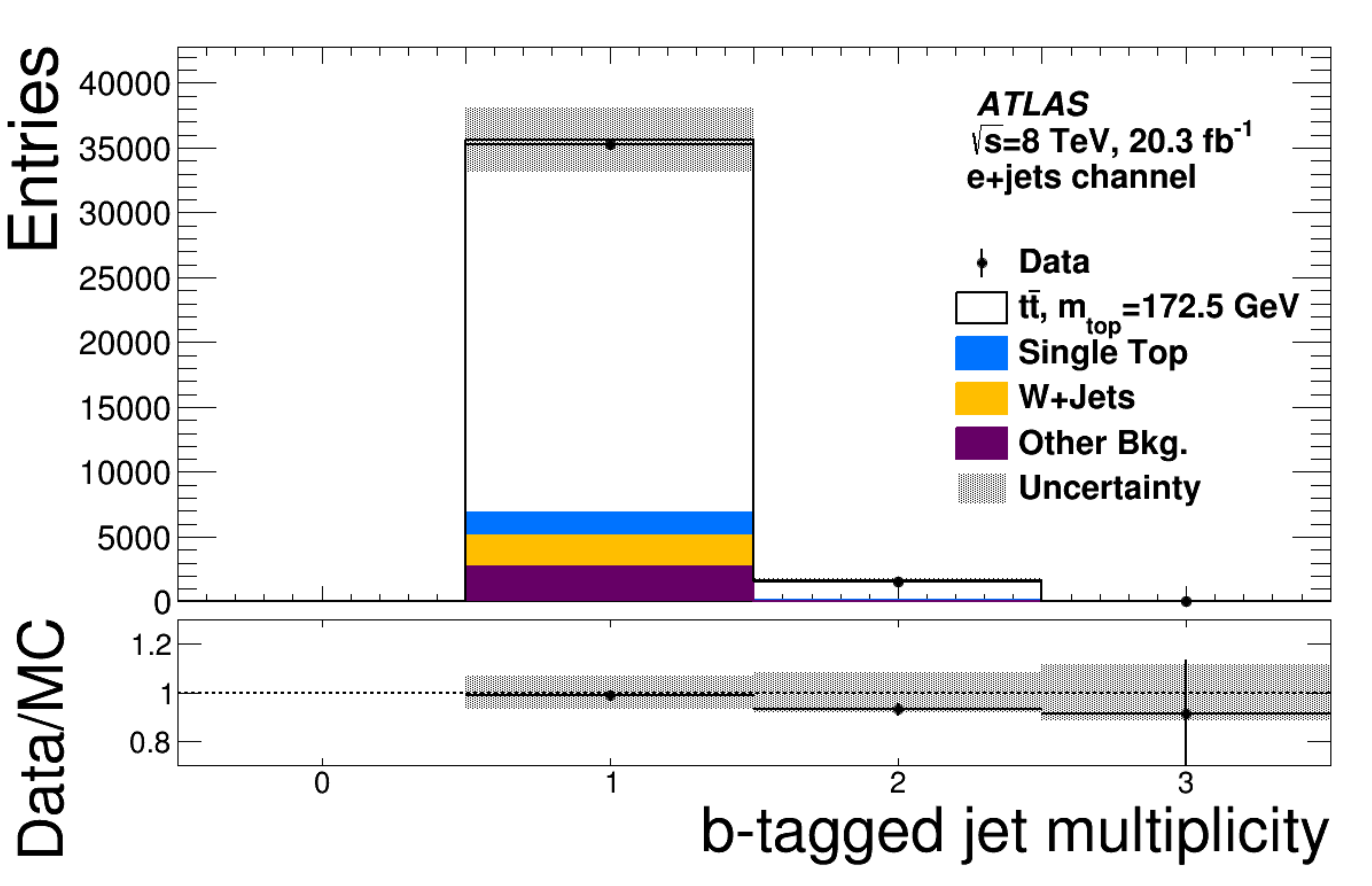}
  }
  \hfill
  \subfloat[]{
    \includegraphics[width=0.45\textwidth]{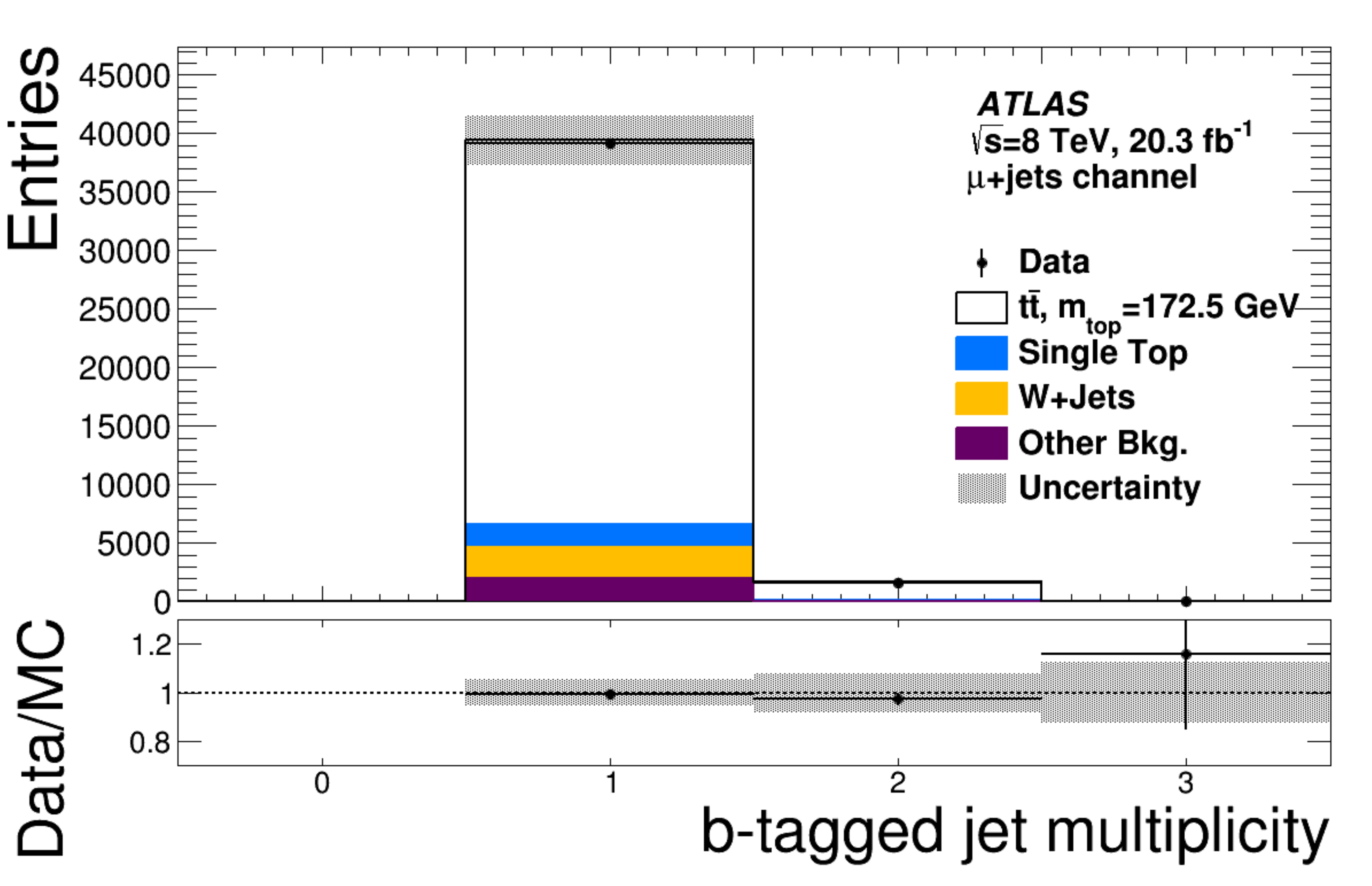}
  }
  
  \vspace{-4mm}
  \subfloat[]{
    \includegraphics[width=0.45\textwidth]{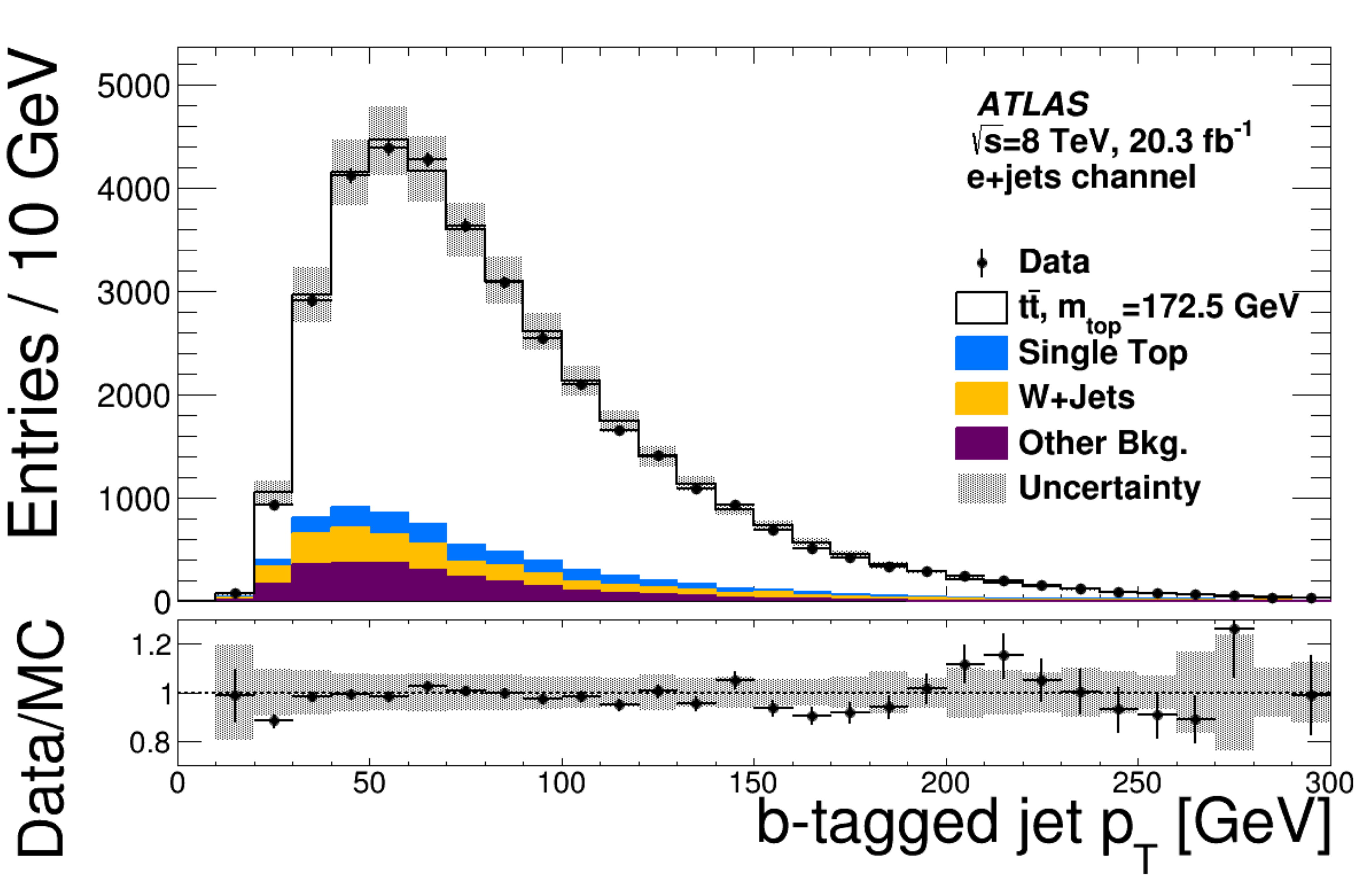}
  }
  \hfill
  \subfloat[]{
    \includegraphics[width=0.45\textwidth]{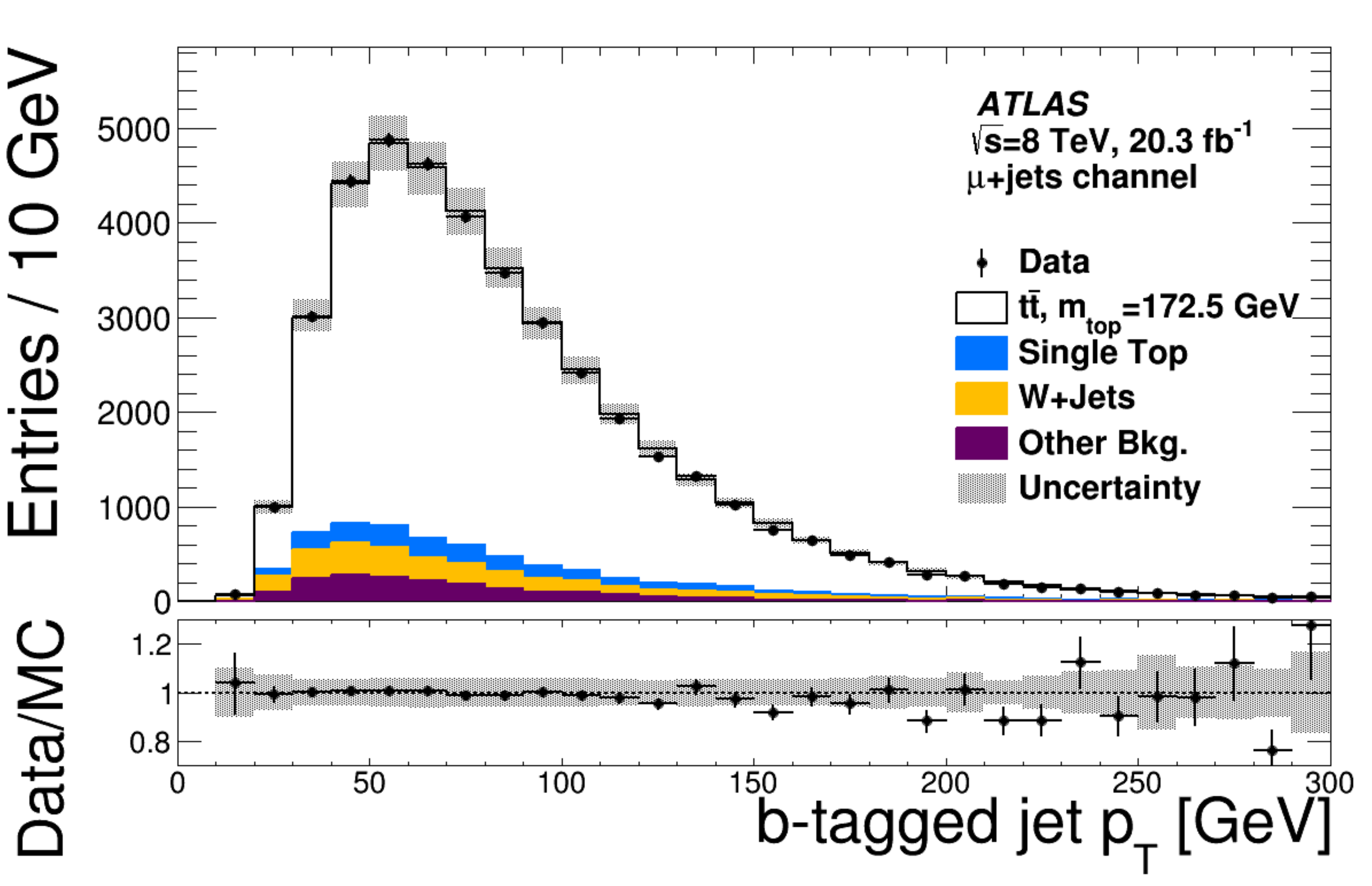}
  }
  
  \vspace{-4mm}
  \subfloat[]{
    \includegraphics[width=0.45\textwidth]{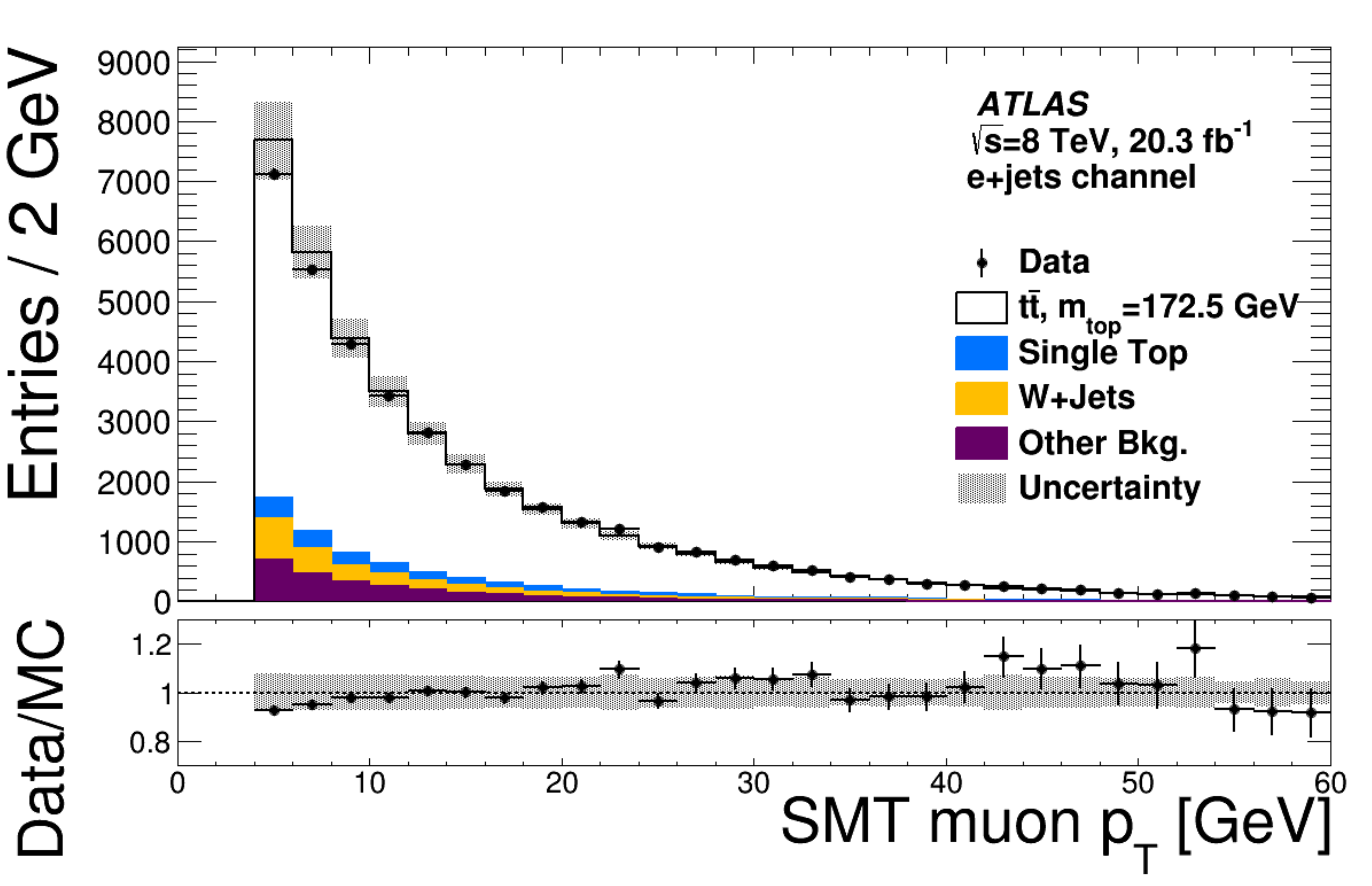}
  }
  \hfill
  \subfloat[]{
    \includegraphics[width=0.45\textwidth]{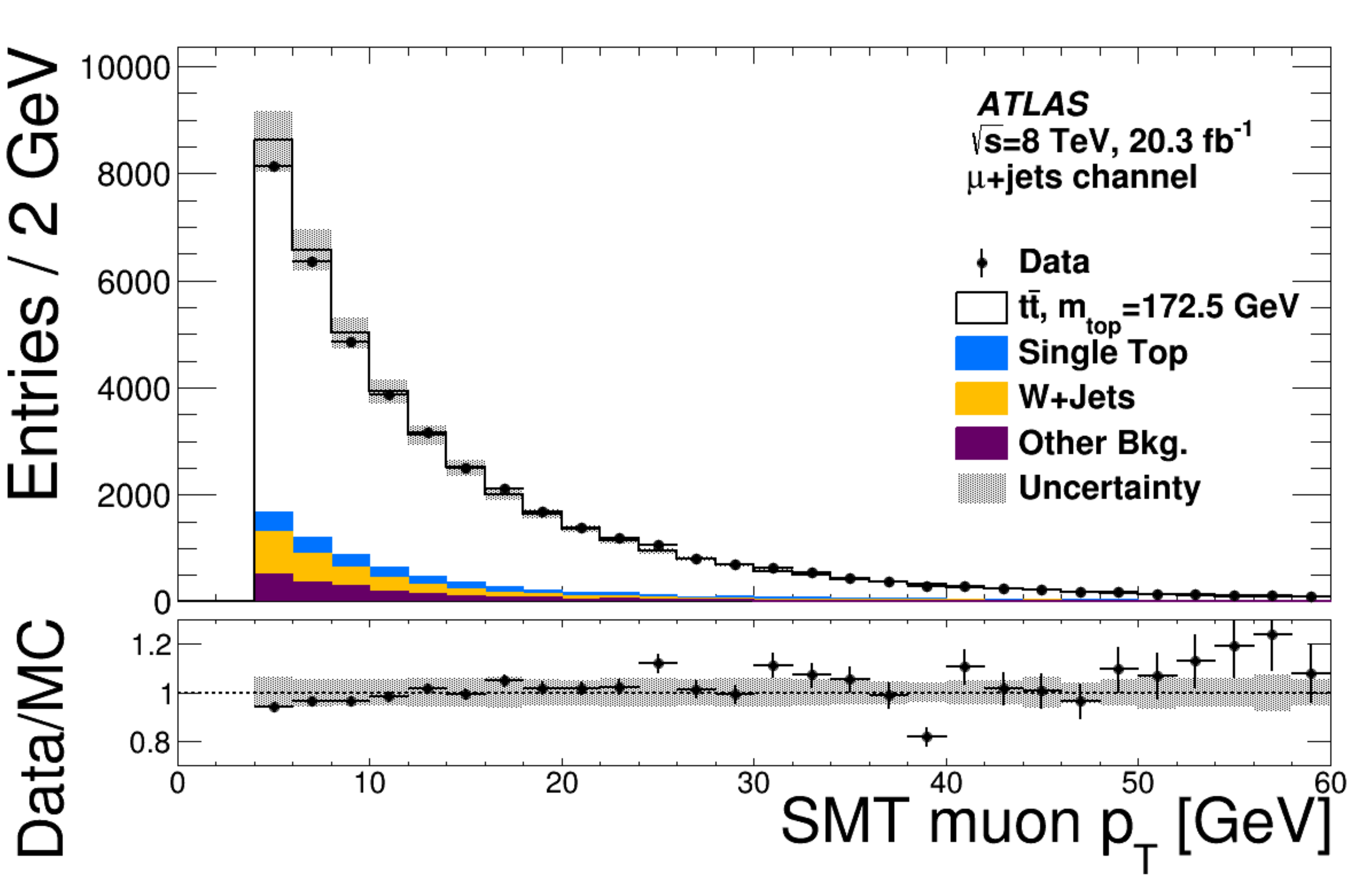}
  }  
  
  \caption{Distributions showing data and simulation. The hashed area represents all experimental systematic uncertainties as well as the $b$-hadron production and hadron-to-muon branching ratio uncertainties. The lower panel of the distributions show the ratio of the data divided by the simulation. The distributions on the left-hand side show the $e$+jets channel while the distributions on the right-hand side show the $\mu$+jets channel. (a) and (b) show the $b$-tagged jet multiplicity, (c) and (d) show the $b$-tagged jet~\pt{} and finally (e) and (f) show the SMT muon~\pt{}.}
  \label{fig:plots}
\end{figure}

The inclusive cross-section, $\sigma_{\ttbar{}}$, is measured in the tagged sample using:

\begin{equation}
\sigma_{\ttbar{}} = \dfrac{N_{\mathrm{data}} - N_{\mathrm{bkg}}}{\int{L\mathrm{d}t} \times \epsilon \times BR},
\label{eq:Yields:crosssection}
\end{equation}

where $N_{\mathrm{data}}$ and $N_{\mathrm{bkg}}$ are the data and background yields respectively, $\epsilon$ is the signal efficiency for single-lepton and dilepton channels (the acceptance for the~\ttbar{} fully hadronic channel is negligible) and $BR=0.543$ is the single-lepton and dilepton total branching ratio derived using a $W\rightarrow \ell\nu$ branching ratio of 0.108 per flavour ($e,\mu,\tau$)~\cite{Agashe:2014kda}. The estimated signal efficiencies and their overall uncertainties are:

\begin{align}
\begin{split}
{\mathrm{\epsilon}}_{\ttbar{}}^{e+\mathrm{jets}} = 0.0109\pm0.0011, \\
{\mathrm{\epsilon}}_{\ttbar{}}^{\mu+\mathrm{jets}} = 0.0124\pm0.0010, \\
{\mathrm{\epsilon}}_{\ttbar{}} = 0.0232\pm0.0020.
\end{split}
\label{Eq:XS:SigEff}
\end{align}

The $e$+jets and $\mu$+jets analyses are combined by summing the event yields for the data, signal and background estimates. Uncertainties are evaluated by recalculating the acceptance and cross-section for each and every individual uncertainty, and then adding the difference from the nominal together in quadrature. The cross-section is measured in the $e$+jets, $\mu$+jets and combined channels and is listed in Table~\ref{Tab:XS:xs_uncert} along with the contributions, in percent, from the systematic uncertainties. The dominant systematic terms come from the jet energy scale, the displaced-vertex $b$-tagging efficiency, the hadron-to-muon branching ratio uncertainties, and MC modelling. The combined cross-section is measured to be:

\begin{equation}
\sigma_{\ttbar{}} = 250 \pm 1 \left({\mathrm{stat.}}\right)  ^{+24}_{-23} \left({\mathrm{syst.}}\right)  \pm 5\left({\mathrm{lumi.}}\right) \pm 4\left({\mathrm{beam}}\right) {\mathrm{pb}}.
\label{Eq:XS:xs_cb}
\end{equation}

The uncertainty of the integrated luminosity is 1.9\%~\cite{CERN-EP-2016-117} and the uncertainty from the knowledge of the LHC beam energy is 1.7\%~\cite{Wenninger:1546734}.

\begin{table}[!htb]
\begin{center}
\begin{tabular}{l|ccc}
 & $e$+jets  & $\mu$+jets & $\ell$+jets\\
$\sigma_{\ttbar{}}$ [pb]  & 248.0 & 251.4 & 249.6\\
\hline
Statistical uncertainty in \%                  & $\pm$0.6 & $\pm$0.6 & $\pm$0.4 \\
\hline
Sources of experimental uncertainty in \% & & & \\
Lepton charge misidentification              & +0.0 $-$0.0 & +0.0 $-$0.0 & +0.0 $-$0.0 \\
Lepton energy resolution         & +1.1 $-$1.0 & +1.0 $-$1.0 & +1.0 $-$1.0 \\
Lepton trigger, reco, identification        & +2.8 $-$2.6 & +2.1 $-$2.0 & +2.1 $-$2.0 \\
Jet energy scale                 & +5.2 $-$5.2 & +4.7 $-$4.6 & +5.0 $-$4.8 \\
Jet energy resolution            & +0.1 $-$0.1 & +0.3 $-$0.3 & +0.1 $-$0.1 \\
Jet reco efficiency              & +0.1 $-$0.1 & +0.1 $-$0.1 & +0.1 $-$0.1 \\
Jet vertex fraction              & +1.0 $-$1.0 & +1.0 $-$1.0 & +1.0 $-$1.0 \\
Fake lepton estimate                        & +4.7 $-$4.7 & +1.0 $-$1.0 & +2.7 $-$2.7 \\
Background normalisation            & +0.2 $-$0.2 & +0.1 $-$0.1 & +0.2 $-$0.2 \\
$W$+jets estimate (statistical)        & +0.0 $-$0.0 & +0.0 $-$0.0 & +0.0 $-$0.0 \\
Single-top production asymmetry            & +0.1 $-$0.0 & +0.1 $-$0.0 & +0.1 $-$0.0 \\
$b$-tagging efficiency       & +2.2 $-$2.1 & +2.2 $-$2.1 & +2.2 $-$2.1 \\
$c$-jet mistag rate     & +0.4 $-$0.4 & +0.4 $-$0.4 & +0.4 $-$0.4 \\
Light-jet mistag rate      & +0.1 $-$0.1 & +0.1 $-$0.1 & +0.1 $-$0.1 \\
SMT reco identification                    & +1.6 $-$1.5 & +1.5 $-$1.5 & +1.5 $-$1.5 \\
SMT momentum imbalance                    & +1.0 $-$1.0 & +1.0 $-$1.0 & +1.0 $-$1.0 \\
SMT light-jet mistag rate                    & +0.4 $-$0.5 & +0.4 $-$0.5 & +0.4 $-$0.5 \\
\hline
Sources of modelling uncertainty in \% & & & \\
Hadron-to-muon branching ratio              & +2.8 $-$2.6 & +2.8 $-$2.5 & +2.8 $-$2.6 \\
$b$-hadron production fractions              & +0.4 $-$0.3 & +0.4 $-$0.4 & +0.4 $-$0.4 \\
Additional radiation             & $\pm$5.3 & $\pm$3.9 & $\pm$4.5 \\
MC generator                     & $\pm$3.0 & $\pm$3.1 & $\pm$3.0 \\
Parton shower                    & $\pm$2.1 & $\pm$1.7 & $\pm$1.9 \\
Parton distribution function          & $\pm$1.1 & $\pm$0.8 & $\pm$0.9 \\
\hline
Total experimental uncertainty              & +8.3 $-$8.1 & +6.2 $-$6.0 & +6.9 $-$6.7 \\
Total modelling  uncertainty                & +7.1 $-$7.0 & +6.0 $-$5.9 & +6.5 $-$6.4 \\
Total systematic uncertainty                & +11 \,\,$-$11 & +8.6 $-$8.4 & +9.4 $-$9.3 \\
\hline
Luminosity uncertainty           & $\pm$1.9 & $\pm$1.9 & $\pm$1.9 \\
LHC beam energy                  & $\pm$1.7 & $\pm$1.7 & $\pm$1.7 \\

\end{tabular}
\caption{\label{Tab:XS:xs_uncert}Measurements of $\sigma_{\ttbar{}}$ for the $e$+jets, $\mu$+jets and combined $\ell$+jets channels, with systematic uncertainties in percent.}
\end{center}
\end{table}

The cross-section measurement presented here is in good agreement with the theoretical prediction and with contemporary results from both ATLAS~\cite{Aad:2014kva,Aad:2015pga} and CMS~\cite{Chatrchyan:2013faa,Khachatryan:2014loa,Khachatryan:2015fwh}.

\FloatBarrier

\section{Measurement of charge asymmetries}
\label{sec:Results}

The data are separated into same- and different-top-like SMT muons, as illustrated in Figure~\ref{fig:AnalysisOutline:SameDiffTop}, by a kinematic likelihood fitter (KLFitter)~\cite{Erdmann:2013rxa}. The KLFitter places Breit--Wigner mass constraints on the top-quark and $W$-boson masses, and permutes reconstructed jets into each possible position in the leading-order parton representation of the~\ttbar{} system. Transfer functions, motivated by detector geometry, are used to map reconstructed jets to partons. For each possible permutation a likelihood and event probability are calculated, and the permutation with the highest event probability is selected. The KLFitter was configured for the $\ell$+jets~\ttbar{} system and optimised for this measurement, with $b$-tagged reconstructed jets being fixed into the KLFitter $b$-jet positions, and allowing at most five reconstructed jets to enter the permutations. The top-quark mass was fixed at the MC mass of $m_{\mathrm{top}}=172.5~\gev{}$. If a reconstructed $b$-tagged jet is mapped to the KLFitter leptonic $b$-jet position then the SMT muon is considered to be same-top-like, whereas if the $b$-tagged jet is mapped to the KLFitter hadronic $b$-jet position then the SMT muon is considered to be different-top-like. In the case of events where both $b$-hadrons decay semileptonically and are both experimentally tagged, one SMT muon is considered same-top-like and the other different-top-like, and both SMT muons contribute to the charge asymmetries.  A misassignment probability of $21\pm1\%$ is achieved. No additional systematic uncertainty is associated with the KLFitter as the algorithm is solely dependent on the four-momenta of the reconstructed objects, which are well described and covered by the existing systematic uncertainties. A consistent KLFitter performance is achieved across all possible charge and same- or different-top configurations, as determined in simulated~\ttbar{} events.

The yield of SMT muons, shown for each charge combination, that are designated as same-top-like is shown in Figure~\ref{fig:fiducial:st_CA} while those designated as different-top-like is shown in Figure~\ref{fig:fiducial:dt_CA}. As stated in Section~\ref{sec:Introduction}, for different-top-like SMT muons, the sign of the $W$-boson lepton has been flipped in order to consistently represent the charge of the $b$-quark at production in both the same- and different-top scenarios.  The observed data are then combined and unfolded to the particle level via: 

\begin{equation}
N^{i} = \frac{1}{\epsilon^{i}}\,\cdot\,\sum_{j} \mathcal{M}^{-1}_{ij}\,\cdot\,f^{j}_{\mathrm{acc}}\,\cdot\,(N^{j}_{\mathrm{data}}-N^{j}_{\mathrm{bkg}}),
\label{eq:fidCPV:unfoldMaster}
\end{equation}

where $i,j=\left\{{\mathrm{N^{++},N^{--},N^{+-},N^{-+}}}\right\}$ and index $i$ runs over the particle level while index $j$ runs over the reconstruction level. $N^{j}_{\mathrm{data}}$ and $N^{j}_{\mathrm{bkg}}$ are the number of SMT muons observed in data and the estimated background, respectively. An acceptance term, $f_{\mathrm{acc}}^{j}$, is applied bin-by-bin to correct for SMT muons that are present at the reconstruction level, but not at the fiducial level. The acceptance term also includes backgrounds within the~\ttbar{} sample itself, such as muons originating from light-flavour, pile-up, $c \rightarrow \mu$, initial- and final-state radiation and dilepton~\ttbar{} events. The response matrix, $\mathcal{M}_{ij}$, is populated exclusively by SMT muons which are matched between the reconstruction and particle level. Finally, an efficiency term, $\epsilon^{i}$, is applied bin-by-bin to the unfolded data to correct for SMT muons that are present at the particle level, but not at the reconstruction level.

The response matrix, $\mathcal{M}_{ij}$, is a discrete $4\times4$ matrix, shown in Table~\ref{tab:fidCPV:cb_response}, where non-zero off-diagonal terms can only occur via charge misidentification or via the misassignment of the same- or different-top SMT muon classification. Charge misidentification was found to be negligible. $\mathcal{M}_{ij}$ is inverted using unregularised matrix inversion, as implemented by the RooUnfold~\cite{Adye:2011gm} program, and is found to show no bias when artificial asymmetries are injected.

\begin{table}[!htb]
\begin{center}
\begin{tabular}{l|cccc}
  & $N^{++}_{j}$ & $N^{--}_{j}$ & $N^{+-}_{j}$ & $N^{-+}_{j}$ \\
\hline\\[0.01ex]
$N^{++}_{i}$ & $0.79 $ & $0.00 $ & $0.00 $ & $0.21 $ \\[1.2ex] 
$N^{--}_{i}$ & $0.00 $ & $0.79 $ & $0.21 $ & $0.00 $ \\[1.2ex] 
$N^{+-}_{i}$ & $0.00 $ & $0.21 $ & $0.79 $ & $0.00 $ \\[1.2ex] 
$N^{-+}_{i}$ & $0.21 $ & $0.00 $ & $0.00 $ & $0.79 $ \\[1.2ex] 
\end{tabular}
\caption{\label{tab:fidCPV:cb_response}Response matrix, $\mathcal{M}_{ij}$. The diagonal elements indicate that the same- or different-top assignment is correct. All non-zero off-diagonal elements come from same- or different-top mistagging. Charge misidentification enters the response matrix at the order of $10^{-5}$. The total uncertainty in all non-zero elements is 0.01.}
\end{center}
\end{table}

The observed charge asymmetries are given in Equations~(\ref{eq:fidCPV:data_Ass}) and~(\ref{eq:fidCPV:data_Aos}) and are found to be compatible with zero: 

\begin{alignat}{4}
  &A^{\mathrm{ss}} &&= -&&0.007 &&\pm 0.006\,\,\,\left({\mathrm{ stat.}}\right) ^{\,\,+0.002}_{\,\,-0.002}\,\,\left({\mathrm{ expt.}}\right) \pm0.005\,\,\,\left({\mathrm{ model}}\right), \label{eq:fidCPV:data_Ass} \\
  &A^{\mathrm{os}} &&= &&0.0041 &&\pm 0.0035\left({\mathrm{ stat.}}\right) ^{\,\,+0.0013}_{\,\,-0.0011}\left({\mathrm{ expt.}}\right) \pm0.0027 \left({\mathrm{ model}}\right). \label{eq:fidCPV:data_Aos}
\end{alignat}

Both the statistical and systematic correlations between $A^{\mathrm{ss}}$ and $A^{\mathrm{os}}$ are estimated to be $\rho_{\mathrm{ss,os}} = -1.0$. 

The systematic uncertainties in each charge asymmetry, shown in Table~\ref{tab:fidCPV:data_syst}, are estimated by keeping the data constant and re-evaluating the acceptance, efficiency, response matrix and background subtraction for each uncertainty component. The largest uncertainty is statistical, which is estimated using 5,000 toy experiments with Poisson-smeared $N_{\mathrm{data}}$ terms in Equation~(\ref{eq:fidCPV:unfoldMaster}). The majority of the systematic uncertainties scale the four charge pair bins uniformly, and their effects cancel when ratios are taken in the construction of $A^{\mathrm{ss}}$ and $A^{\mathrm{os}}$. Other systematic uncertainties have some charge dependence (such as the single top background) or affect the performance of KLFitter (such as the additional radiation), as such these uncertainties are more prominent. The additional radiation and PDF uncertainties form the largest modelling uncertainty, whilst the jet energy scale and lepton energy resolution are significant experimental uncertainties.

The MC simulation predictions in the fiducial region are obtained using the nominal~\ttbar{} simulation, which contains no sources of CP violation, and the uncertainties are estimated by using the modelling uncertainties described in Section~\ref{sec:syst:model}. The MC simulation predictions are shown in Equations~(\ref{eq:fidCPV:sim_Ass}) and~(\ref{eq:fidCPV:sim_Aos}).

\begin{alignat}{4}
&A^{\mathrm{ss}}_{\mathrm{sim}} &&= &&0.0005 &&\pm 0.0016 \label{eq:fidCPV:sim_Ass} \\
&A^{\mathrm{os}}_{\mathrm{sim}} &&= -&&0.0003 &&\pm 0.0009 \label{eq:fidCPV:sim_Aos}
\end{alignat}

The MC simulation is found to be in good agreement with the data. The uncertainties shown come from MC statistics, which are of the same size as those found from propagating the signal modelling uncertainties. Both are compatible with zero and with the SM predictions of $\lvert A^{\mathrm{ss}}\rvert \left(\lvert A^{\mathrm{os}}\rvert \right) < 10^{-4}$~\cite{Gedalia:2012sx}.

\begin{table}[!htb]
\begin{center}

\begin{tabular}{l|lllr}
 & \multicolumn{2}{c}{$A^{\mathrm{ss}}\left(10^{-2}\right)$} & \multicolumn{2}{c}{$A^{\mathrm{os}}\left(10^{-2}\right)$}  \\
Measured value                       & \multicolumn{2}{c}{$-$0.7} & \multicolumn{2}{c}{0.41}  \\
\hline
Statistical   uncertainty                     & \multicolumn{2}{c}{$\pm$0.6} & \multicolumn{2}{c}{$\pm$0.35} \\
\hline
Sources of experimental uncertainty & & & & \\
Lepton charge misidentification             & +0.002 &$-$0.002 & +0.001 &$-$0.001 \\
Lepton energy resolution                     & +0.09 &$-$0.11 & +0.07 &$-$0.06 \\
Lepton trigger, reco, identification         & +0.004 &$-$0.004 & +0.002 &$-$0.002 \\
Jet energy scale                 & +0.10 &$-$0.14 & +0.08 &$-$0.06 \\
Jet energy resolution            & +0.019 &$-$0.019 & +0.009 &$-$0.009 \\
Jet reco efficiency              & +0.010 &$-$0.010 & +0.006 &$-$0.006 \\
Jet vertex fraction              & +0.09 &$-$0.09 & +0.05 &$-$0.05 \\
Fake lepton estimate                        & +0.05 &$-$0.05 & +0.025 &$-$0.025 \\
Background normalisation            & +0.002 &$-$0.002 & +0.001 &$-$0.001 \\
$W$+jets estimate (statistical)        & +0.003 &$-$0.002 & +0.001 &$-$0.002 \\
Single-top production asymmetry          & +0.016 &$-$0.002 & +0.001 &$-$0.009 \\
$b$-tagging efficiency        & +0.008 &$-$0.008 & +0.004 &$-$0.004 \\
$c$-jet mistag rate     & +0.020 &$-$0.020 & +0.013 &$-$0.013 \\
Light-jet mistag rate    & +0.022 &$-$0.023 & +0.013 &$-$0.012 \\
SMT  reco identification                  & +0.004 &$-$0.004 & +0.004 &$-$0.004 \\
SMT   momentum imbalance                  & +0.06 &$-$0.06 & +0.04 &$-$0.035 \\
SMT  light-jet mistag rate                   & +0.010 &$-$0.009 & +0.005 &$-$0.005 \\
\hline
Sources of modelling uncertainty & & & &  \\
Hadron-to-muon branching ratio            & +0.04 &$-$0.05 & +0.026 &$-$0.022 \\
$b$-hadron production              & +0.013 &$-$0.008 & +0.003 &$-$0.008 \\
Additional radiation             & \multicolumn{2}{c}{$\pm$0.4} & \multicolumn{2}{c}{$\pm$0.23} \\
MC generator                     & \multicolumn{2}{c}{\,\,\,$\pm$0.05} & \multicolumn{2}{c}{\,\,\,$\pm$0.025} \\
Parton shower                    & \multicolumn{2}{c}{\,\,\,$\pm$0.04} & \multicolumn{2}{c}{\,\,\,$\pm$0.017} \\
Parton distribution function          & \multicolumn{2}{c}{\,\,\,$\pm$0.22} & \multicolumn{2}{c}{$\pm$0.13} \\
\hline
Total experimental uncertainty                & +0.19 &$-$0.22 & +0.13 &$-$0.11 \\
Total modelling  uncertainty                  & +0.5 &$-$0.5 & +0.27 &$-$0.27 \\
Total systematic uncertainty                & +0.5 &$-$0.5 & +0.30 &$-$0.29 \\
\end{tabular}
\caption{\label{tab:fidCPV:data_syst}Measurements $A^{\mathrm{ss}}$ and $A^{\mathrm{os}}$, in units of $10^{-2}$, and breakdown of absolute uncertainties. }
\end{center}
\end{table}

\begin{figure}[!htb]
  \subfloat[$e$+jets channel.]{
    \includegraphics[width=0.45\textwidth]{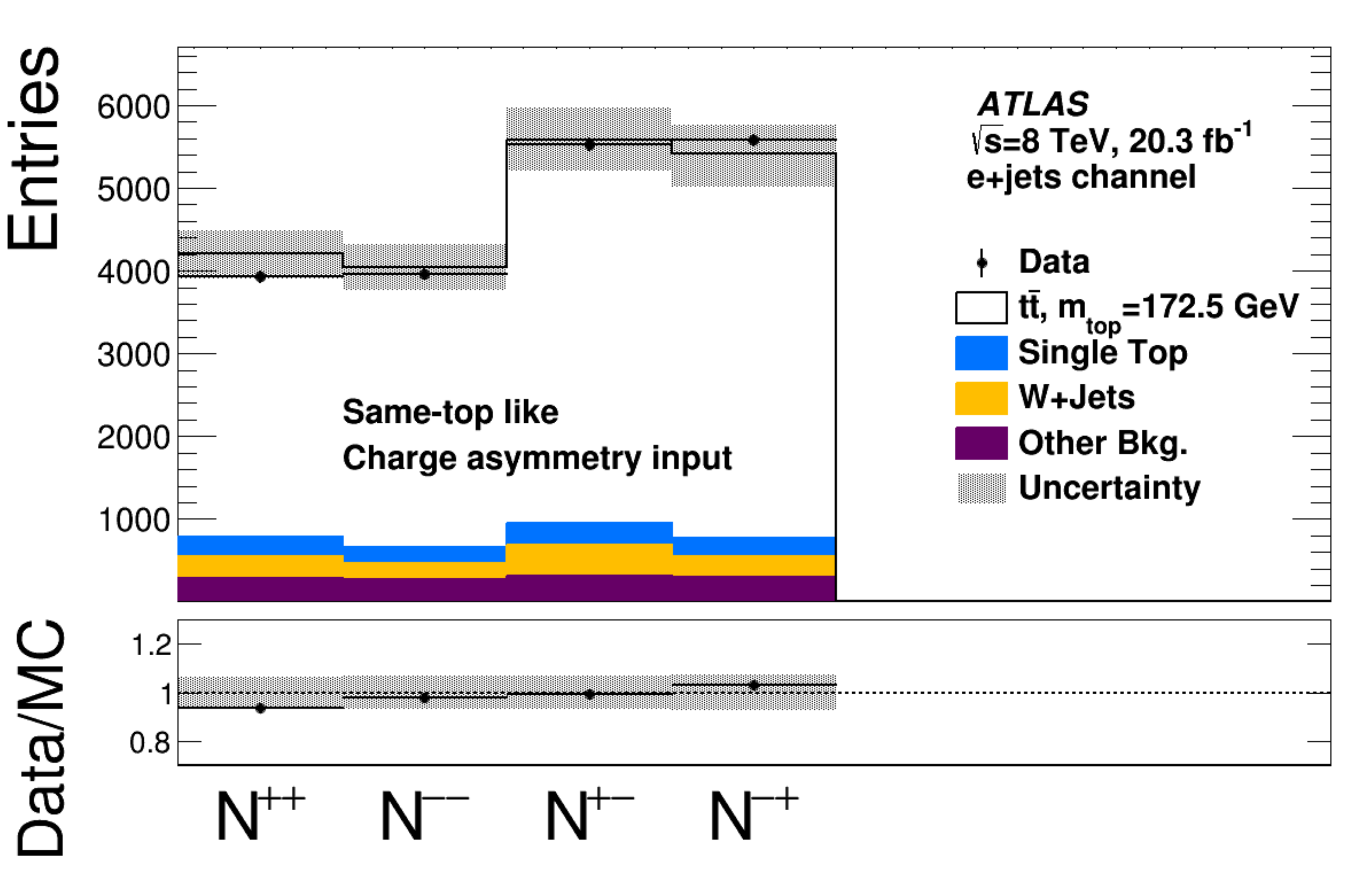}
  }
  \hfill
  \subfloat[$\mu$+jets channel.]{
    \includegraphics[width=0.45\textwidth]{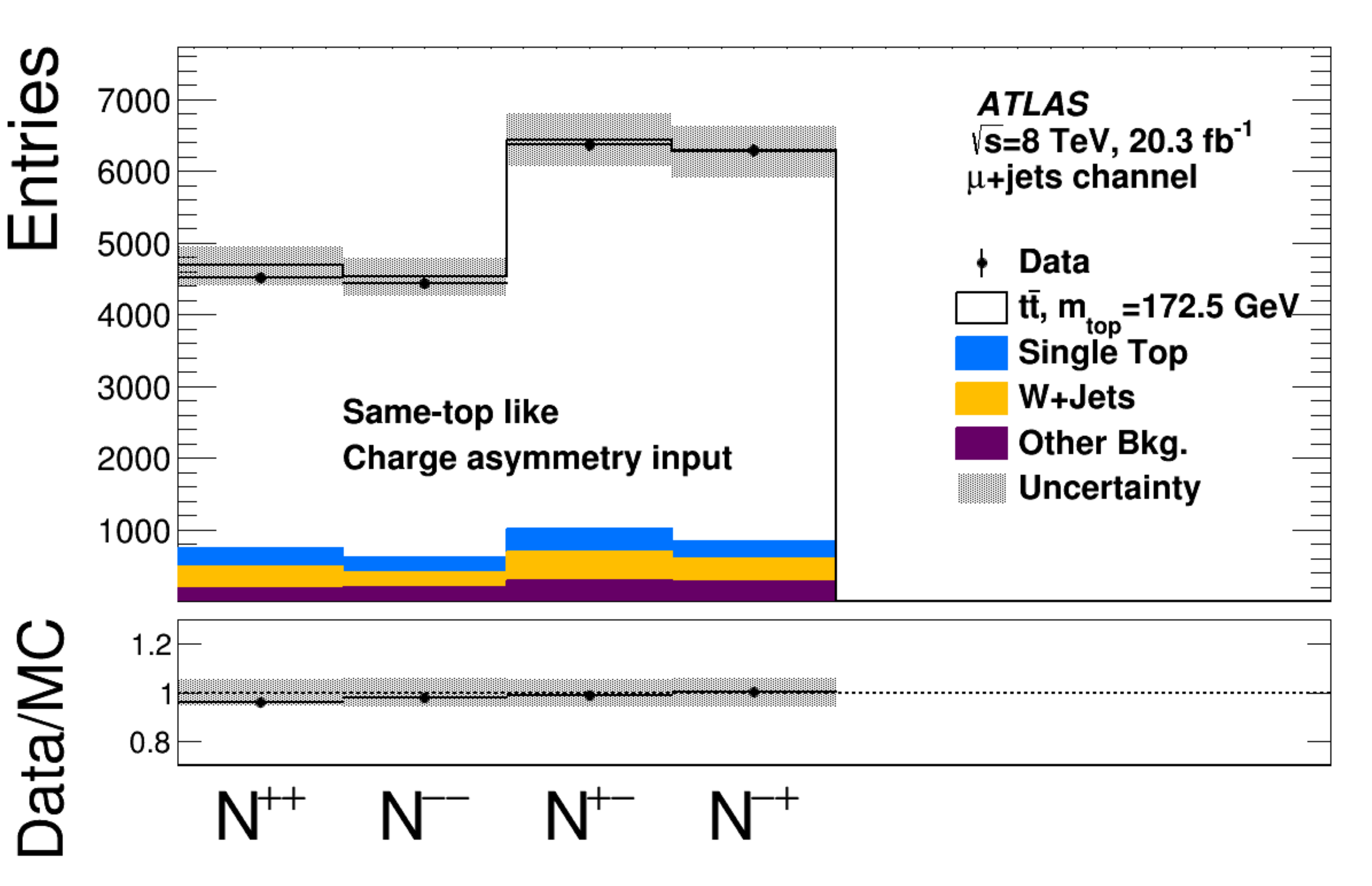}
  }
  \caption{Same-top-like charge-pairings distributions. The hashed area represents all experimental systematic uncertainties as well as the $b$-hadron production and hadron-to-muon branching ratio uncertainties. The lower panel of the distributions show the ratio of the data divided by the simulation. (a) shows the $e$+jets channel while (b) shows the $\mu$+jets channel.}
  \label{fig:fiducial:st_CA}
\end{figure}

\begin{figure}[!htb]
  \subfloat[$e$+jets channel.]{
    \includegraphics[width=0.45\textwidth]{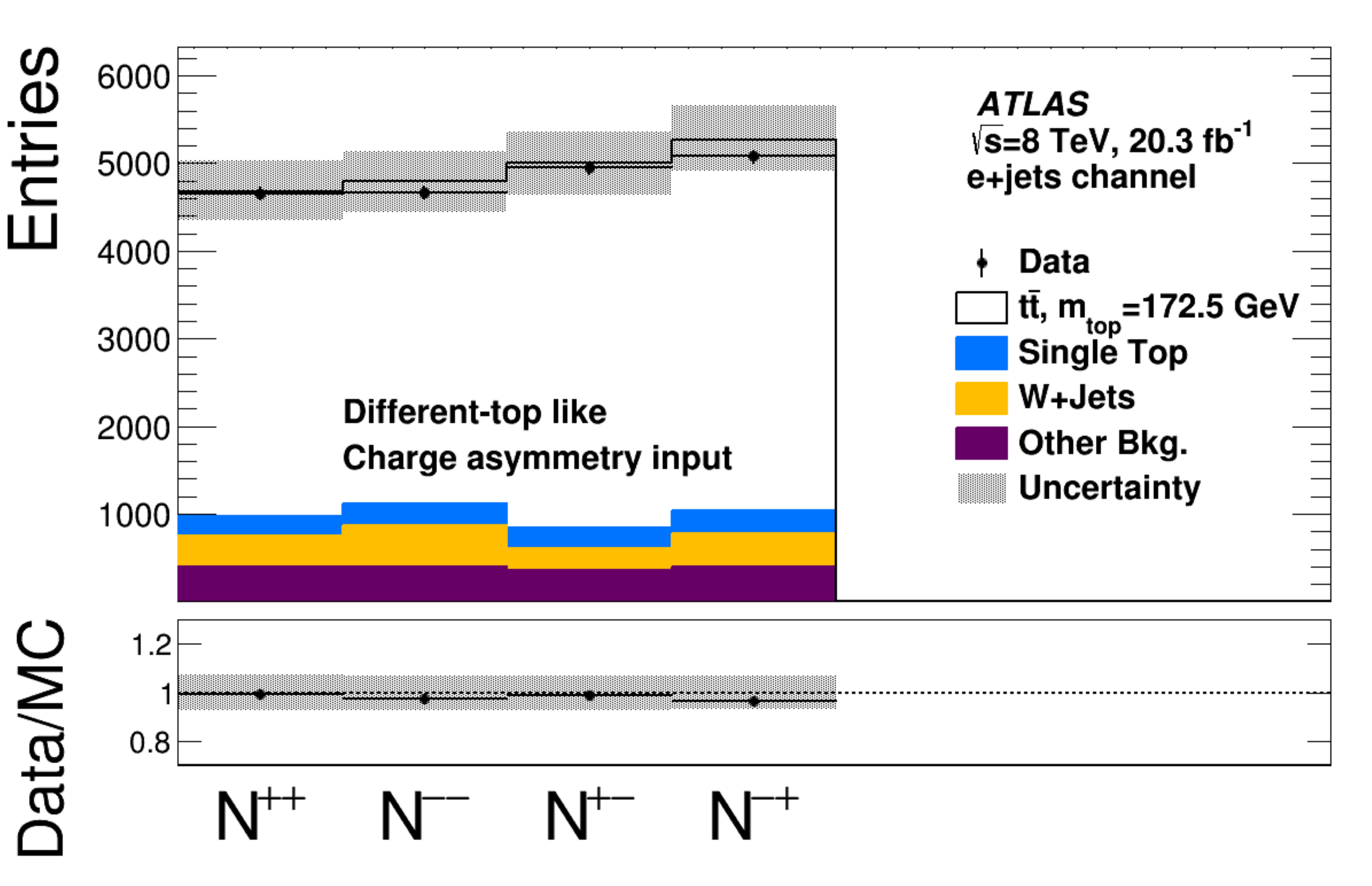}
  }
  \hfill
  \subfloat[$\mu$+jets channel.]{
    \includegraphics[width=0.45\textwidth]{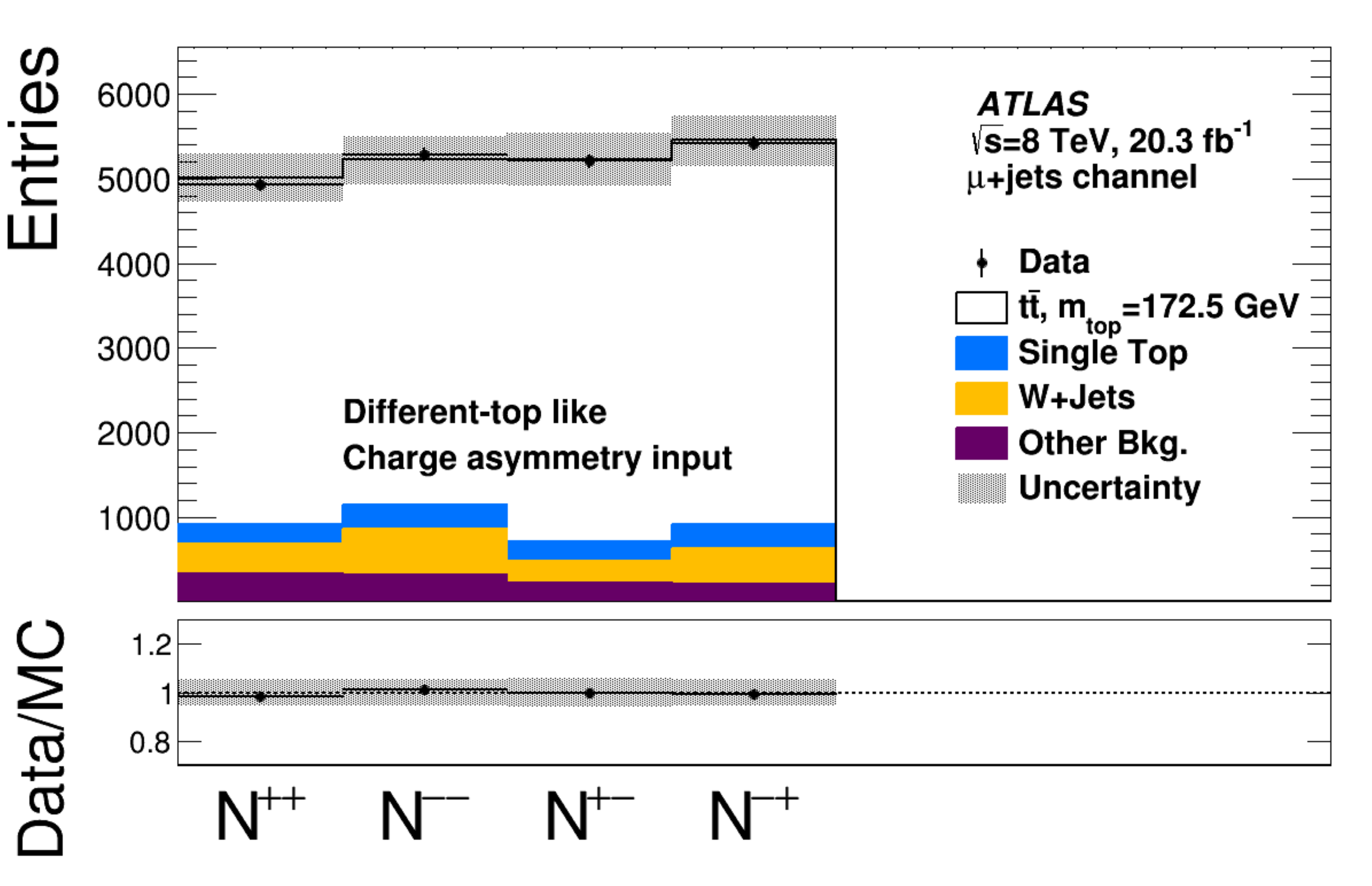}
  }
  \caption{Different-top-like charge-pairings distributions. The hashed area represents all experimental systematic uncertainties as well as the $b$-hadron production and hadron-to-muon branching ratio uncertainties. The lower panel of the distributions show the ratio of the data divided by the simulation. (a) shows the $e$+jets channel while (b) shows the $\mu$+jets channel.}
  \label{fig:fiducial:dt_CA}
\end{figure}

\section{Interpretation of the charge asymmetries}
\label{sec:interpetationCPV}

The decay-chain fractions are obtained from simulation at the particle level, and are detailed in Table~\ref{tab:app_inter:ri}.  They can be used in conjunction with the observed charge asymmetries in order to extract the various CP asymmetries. The largest uncertainties in the decay-chain fractions come from the hadron-to-muon branching ratio and the parton shower. There are two observed charge asymmetries and five CP asymmetries, leading to an underconstrained system. Following the suggestion of Ref.~\cite{Gedalia:2012sx}, each CP asymmetry in this interpretation is considered in turn whilst setting the other four CP asymmetries to zero. Furthermore, following the convention of Refs.~\cite{Grossman:2009} and~\cite{Kagan:2009}, in the case of zero direct CP violation, $A^{b\ell}_{\mathrm{mix}} \equiv A^{bc}_{\mathrm{mix}}$ and is hereafter referred to as $A^{b}_{\mathrm{mix}}$.  For any CP asymmetries appearing in both charge asymmetries, the tighter of the constraints is taken. For $A^{b}_{\mathrm{mix}}$ the tighter measurement comes solely from the $A^{\mathrm{ss}}$ charge asymmetry. This technique results in the following CP asymmetries:

\begin{alignat}{5}
& A^{b}_{\mathrm{mix}} && = \dfrac{A^{\mathrm{ss}}}{r_{b} + r_{c\overline{c}}} && = -&&0.025 &&\pm 0.021 \left({\mathrm{stat.}}\right) \pm0.008\left({\mathrm{expt.}}\right) \pm0.017 \left({\mathrm{model}}\right), \label{Eq:int:AbMix} \\[1ex]
& A^{b\ell}_{\mathrm{dir}} &&= \dfrac{A^{\mathrm{os}}}{\widetilde{r}_{b}} && =  &&0.005 &&\pm 0.004 \left({\mathrm{stat.}}\right) \pm0.001 \left({\mathrm{expt.}}\right) \pm0.003 \left({\mathrm{model}}\right),\label{Eq:int:AblDir} \\[1ex]
& A^{c\ell}_{\mathrm{dir}} &&= \dfrac{-A^{\mathrm{ss}}}{r_{c} + r_{c\overline{c}}} && = &&0.009 &&\pm 0.007 \left({\mathrm{stat.}}\right) \pm0.003 \left({\mathrm{expt.}}\right) \pm0.006 \left({\mathrm{model}}\right), \label{Eq:int:AclDir} \\[1ex]
& A^{bc}_{\mathrm{dir}} &&= \dfrac{A^{\mathrm{ss}}}{r_{c}} && = -&&0.010 &&\pm 0.008 \left({\mathrm{stat.}}\right) \pm0.003 \left({\mathrm{expt.}}\right) \pm0.007 \left({\mathrm{model}}\right).  \label{Eq:int:AbcDir}
\end{alignat}

with the systematic uncertainties shown in Table~\ref{Tab:int:ParamUncert}. The predictions of the MC simulation are:

\begin{alignat}{4}
&A^{b}_{\mathrm{mix,sim}} &&= &&0.002 &&\pm 0.005, \label{eq:fidCPV:sim_AbMix} \\
&A^{b\ell}_{\mathrm{dir,sim}} &&= &&0.000 &&\pm 0.001, \label{eq:fidCPV:sim_AblDir} \\
&A^{c\ell}_{\mathrm{dir,sim}} &&= -&&0.0006 &&\pm 0.0019, \label{eq:fidCPV:sim_AclDir} \\
&A^{bc}_{\mathrm{dir,sim}} &&= &&0.0007 &&\pm 0.0022, \label{eq:fidCPV:sim_AbcDir}
\end{alignat}

and are found to be in good agreement with the data. The uncertainties shown come from MC statistics, which are of the same size as those found from propagating the signal modelling uncertainties. Both the data and the MC simulation are compatible with zero and with the SM predictions, as shown in Table~\ref{Tab:inter:CP_comp}.

\begin{table}[!htb]
\begin{center}
\begin{adjustbox}{max width=\textwidth}
\begin{tabular}{l|cccccc}
                                 & $r_{b}$ & $r_{c}$ & $r_{c\overline{c}}$ & $\widetilde{r}_{b}$ & $\widetilde{r}_{c}$ & $\widetilde{r}_{c\overline{c}}$ \\
\hline
Nominal                          & 0.200 & 0.715 & 0.085 & 0.882 & 0.069 & 0.048 \\
\hline
Relative uncertainty in \%  &  &  &  &  &  &  \\
Hadron-to-muon branching ratio             & +3.8 $-$3.2 & +2.9 $-$2.3 & +23 \,\,$-$30   & +1.6 $-$1.3 & +3.3 $-$3.3 & +25 $-$31 \\
$b$-hadron production                      & +1.8 $-$1.8 & +0.5 $-$0.5 & +0.3 $-$0.3 & +0.2 $-$0.2 & +1.9 $-$1.9 & +0.2 $-$0.2 \\
Additional radiation                       & $\pm$2.4    & $\pm$0.6    & $\pm$0.4    & $\pm$0.1    & $\pm$0.9    & $\pm$1.1 \\
MC generator                               & $\pm$0.2    & $\pm$0.1    & $\pm$0.1    & $\pm$0.1    & $\pm$0.5    & $\pm$0.7 \\
Parton shower                              & $\pm$6.8    & $\pm$2.2    & $\pm$2.6    & $\pm$0.6    & $\pm$12     & $\pm$6.1 \\
Parton distribution function               & $\pm$0.1    & $\pm$0.1    & $\pm$0.9    & $\pm$0.0    & $\pm$0.3    & $\pm$0.2 \\
\hline
Total uncertainty                          & +8.4 $-$8.1 & +3.7 $-$3.3 & +23 \,\,$-$30   & +1.7 $-$1.4 & +13 $-$13 & +25 $-$31 \\
\end{tabular}
\end{adjustbox}
\caption{\label{tab:app_inter:ri}Decay-chain fractions obtained from MC simulation at the particle level. Uncertainties are in percent.}
\end{center}
\end{table}

\begin{table}[!htb]
\begin{center}
\begin{adjustbox}{max width=\textwidth}
\begin{tabular}{l|ll|llllll}
 & \multicolumn{2}{c|}{$A^{b}_{\mathrm{mix}}\left(10^{-2}\right)$} & \multicolumn{2}{c}{$A^{b\ell}_{\mathrm{dir}}\left(10^{-2}\right)$} & \multicolumn{2}{c}{$A^{c\ell}_{\mathrm{dir}}\left(10^{-2}\right)$} & \multicolumn{2}{c}{$A^{bc}_{\mathrm{dir}}\left(10^{-2}\right)$} \\
Measured value                       & \multicolumn{2}{c|}{$-$2.5} & \multicolumn{2}{c}{0.5} & \multicolumn{2}{c}{0.9} & \multicolumn{2}{c}{$-$1.0} \\
\hline
Statistical  uncertainty                    & \multicolumn{2}{c|}{$\pm$ 2.1} & \multicolumn{2}{c}{$\pm$ 0.4} & \multicolumn{2}{c}{$\pm$ 0.7} & \multicolumn{2}{c}{$\pm$ 0.8} \\
\hline
Sources of experimental uncertainty & & & & & & & &\\
Lepton charge misidentification             & +0.008 &$-$0.007 & +0.001 &$-$0.002 & +0.002 &$-$0.003 & +0.003 &$-$0.003 \\
Lepton energy resolution         & +0.33 &$-$0.39 & +0.07 &$-$0.06 & +0.14 &$-$0.12 & +0.13 &$-$0.15 \\
Lepton trigger, reco, identification       & +0.016 &$-$0.015 & +0.003 &$-$0.003 & +0.005 &$-$0.006 & +0.006 &$-$0.006 \\
Jet energy scale                 & +0.4 &$-$0.5 & +0.09 &$-$0.07 & +0.17 &$-$0.13 & +0.15 &$-$0.19 \\
Jet energy resolution            & +0.07 &$-$0.07 & +0.011 &$-$0.011 & +0.024 &$-$0.024 & +0.027 &$-$0.027 \\
Jet reco efficiency              & +0.034 &$-$0.034 & +0.006 &$-$0.006 & +0.012 &$-$0.012 & +0.014 &$-$0.014 \\
Jet vertex fraction              & +0.33 &$-$0.33 & +0.06 &$-$0.06 & +0.12 &$-$0.12 & +0.13 &$-$0.13 \\
Fake lepton  estimate                       & +0.18 &$-$0.19 & +0.029 &$-$0.029 & +0.07 &$-$0.07 & +0.07 &$-$0.08 \\
Background normalisation            & +0.008 &$-$0.009 & +0.001 &$-$0.001 & +0.003 &$-$0.003 & +0.003 &$-$0.003 \\
$W$+jets estimate (statistical)       & +0.009 &$-$0.008 & +0.002 &$-$0.002 & +0.003 &$-$0.003 & +0.004 &$-$0.003 \\
Single-top production asymmetry           & +0.06 &$-$0.01 & +0.002 &$-$0.011 & +0.002 &$-$0.020 & +0.022 &$-$0.003 \\
$b$-tagging efficiency        & +0.028 &$-$0.028 & +0.005 &$-$0.005 & +0.010 &$-$0.010 & +0.011 &$-$0.011 \\
$c$-jet mistag rate     & +0.07 &$-$0.07 & +0.015 &$-$0.015 & +0.025 &$-$0.026 & +0.029 &$-$0.027 \\
Light-jet mistag rate     & +0.08 &$-$0.08 & +0.014 &$-$0.014 & +0.028 &$-$0.028 & +0.031 &$-$0.032 \\
SMT reco identification                     & +0.013 &$-$0.012 & +0.004 &$-$0.004 & +0.004 &$-$0.005 & +0.005 &$-$0.005 \\
SMT momentum imbalance                     & +0.21 &$-$0.22 & +0.04 &$-$0.04 & +0.08 &$-$0.08 & +0.09 &$-$0.09 \\
SMT light-jet mistag rate                   & +0.035 &$-$0.031 & +0.005 &$-$0.006 & +0.011 &$-$0.012 & +0.014 &$-$0.012 \\
\hline
Sources of modelling uncertainty & & & & & & & &\\
Hadron-to-muon branching ratio              & +0.25 &$-$0.36 & +0.023 &$-$0.020 & +0.06 &$-$0.05 & +0.04 &$-$0.04 \\
$b$-hadron production fractions            & +0.031 &$-$0.021 & +0.004 &$-$0.010 & +0.013 &$-$0.020 & +0.022 &$-$0.015 \\
Additional radiation             & \multicolumn{2}{c|}{$\pm$1.4} & \multicolumn{2}{c}{$\pm$0.26} & \multicolumn{2}{c}{$\pm$0.6} & \multicolumn{2}{c}{$\pm$0.6} \\
MC generator                     & \multicolumn{2}{c|}{\,\,\,$\pm$0.17} & \multicolumn{2}{c}{\,\,\,$\pm$0.029} & \multicolumn{2}{c}{\,\,\,$\pm$0.07} & \multicolumn{2}{c}{\,\,\,$\pm$0.08} \\
Parton shower                    & \multicolumn{2}{c|}{\,\,\,$\pm$0.08} & \multicolumn{2}{c}{\,\,\,$\pm$0.021} & \multicolumn{2}{c}{\,\,\,$\pm$0.06} & \multicolumn{2}{c}{\,\,\,$\pm$0.07} \\
Parton distribution function          & \multicolumn{2}{c|}{$\pm$0.8} & \multicolumn{2}{c}{$\pm$0.15} & \multicolumn{2}{c}{\,\,\,$\pm$0.29} & \multicolumn{2}{c}{\,\,\,$\pm$0.32} \\
\hline
Total experimental uncertainty               & +0.7 &$-$0.8 & +0.14 &$-$0.12 & +0.27 &$-$0.24 & +0.27 &$-$0.31 \\
Total modelling  uncertainty                  & +1.6 &$-$1.7 & +0.30 &$-$0.30 & +0.6 &$-$0.6 & +0.7 &$-$0.7 \\
Total systematic uncertainty                & +1.8 &$-$1.8 & +0.34 &$-$0.33 & +0.7 &$-$0.6 & +0.7 &$-$0.7 \\
\end{tabular}
\end{adjustbox}
\caption{\label{Tab:int:ParamUncert}Measurements of $A^{b}_{\mathrm{mix}}$, $A^{b\ell}_{\mathrm{dir}}$, $A^{c\ell}_{\mathrm{dir}}$ and $A^{bc}_{\mathrm{dir}}$, in units of $10^{-2}$, and breakdown of absolute uncertainties. }
\end{center}
\end{table}

The anomalous dimuon asymmetry observed by the D0 experiment may be interpreted in terms of both mixing and direct CP asymmetries separately, as discussed in Ref.~\cite{Descotes:2015}. If that asymmetry is considered to be caused only by CP violation in mixing, it is calculated that $A^{b}_{\mathrm{mix}} \approx -0.008 \pm 0.003$ would be required to explain such a result. This is at odds with the SM but is not currently excluded by existing measurements of the flavour specific mixing asymmetries $a^{d}_{\mathrm{sl}}$ and $a^{s}_{\mathrm{sl}}$ (which combine as described in Ref.~\cite{Gedalia:2012sx} with $B^{0}_{s,d}$ fragmentation fractions $\left(f_{d},f_{s}\right)$~\cite{Amhis:2014hma} to build $A^{b}_{\mathrm{mix}}$). The world averages for $a^{d}_{\mathrm{sl}}$ and $a^{s}_{\mathrm{sl}}$ are currently at a precision of the order of $10^{-3}$~\cite{Amhis:2014hma}. The result presented here for $A^{b}_{\mathrm{mix}}$ does not have the precision to shed more light in this area. Alternatively, the dimuon asymmetry may be considered to be caused exclusively by CP violation in direct decays. This interpretation requires that either the true  $A^{b\ell}_{\mathrm{dir}} \approx (0.003 \pm 0.001)$ or the true $A^{c\ell}_{\mathrm{dir}} \approx (0.009 \pm 0.003)$, whereas the SM predictions for these parameters are of the order $ \lvert A^{b\ell}_{\mathrm{dir}}\rvert < 10^{-7}$ and $\lvert A^{c\ell}_{\mathrm{dir}}\rvert < 10^{-11}$~\cite{Gedalia:2012sx}~\cite{Descotes:2015}.

The results presented here for direct CP violation and CP violation in mixing are compatible with both the SM predictions and the dimuon asymmetry observed by the D0 measurement, within $1 \sigma$.

For comparison with existing experimental limits on these parameters the discussion presented in Ref.~\cite{Descotes:2015} may be considered, where it is stated that only limits from exclusive channels on some direct asymmetries presently exist.  These limits are extrapolated (extrap.) to inclusive limits by considering uncertainties on the relevant branching ratios. They are evaluated in decay modes insensitive to the other direct CP asymmetries respectively and therefore make no assumptions on their values.  The limits set at the $2\sigma$ level are $A^{b\ell}_{\mathrm{dir}}\left(\textrm{extrap.}\right) \leq 0.012$ and $A^{c\ell}_{\mathrm{dir}}\left(\textrm{extrap.}\right) \leq 0.06$.  No limits are set on $A^{bc}_{\mathrm{dir}}$.

A full comparison of the experimental results with the SM predictions and with existing experimental limits may be found in Table~\ref{Tab:inter:CP_comp}. In particular, the $2 \sigma$ limits inferred by this analysis are stronger than the existing indirect limit on $A^{c\ell}_{\mathrm{dir}}$, and equivalent to the existing indirect limit on $A^{b\ell}_{\mathrm{dir}}$.  Moreover, this is the first direct experimental limit on either of these direct CP asymmetries, and also the first direct experimental limit on $A^{bc}_{\mathrm{dir}}$.

\begin{table}[!htb]
\begin{center}
\begin{adjustbox}{max width=\textwidth}
\begin{tabular}{l|rlrl|clrlr}
                                 & \multicolumn{2}{c}{Data $\left(10^{-2}\right)$} & \multicolumn{2}{c|}{MC $\left(10^{-2}\right)$} & \multicolumn{3}{c}{Existing limits ($2 \sigma$) $\left(10^{-2}\right)$} & \multicolumn{2}{c}{SM prediction $\left(10^{-2}\right)$} \\
\hline
$A^{\mathrm{ss}}$                 & $-$0.7 &$\pm$ 0.8  & 0.05 & $\pm$ 0.23 & \multicolumn{3}{c}{-} & $< 10^{-2}$&~\cite{Gedalia:2012sx} \\
$A^{\mathrm{os}}$                 & 0.4    &$\pm$ 0.5  & $-$0.03    & $\pm$ 0.13 & \multicolumn{3}{c}{-} & $< 10^{-2}$&~\cite{Gedalia:2012sx} \\
$A^{b}_{\mathrm{mix}}$            & $-$2.5 &$\pm$ 2.8 & 0.2    & $\pm$ 0.7 & $\quad \quad \quad$ &$< 0.1$              & ~\cite{Amhis:2014hma} & $< 10^{-3}$&~\cite{Artuso:2015swg}~\cite{Amhis:2014hma} \\
$A^{b\ell}_{\mathrm{dir}}$        & 0.5    &$\pm$ 0.5 & $-$0.03 & $\pm$ 0.14 & &$< 1.2$                & ~\cite{Descotes:2015} & $< 10^{-5}$&~\cite{Gedalia:2012sx}~\cite{Descotes:2015} \\
$A^{c\ell}_{\mathrm{dir}}$        & 1.0    &$\pm$ 1.0 & $-$0.06 & $\pm$ 0.25 & &$< 6.0$                & ~\cite{Descotes:2015} & $< 10^{-9}$&~\cite{Gedalia:2012sx}~\cite{Descotes:2015} \\
$A^{bc}_{\mathrm{dir}}$           & $-$1.0    &$\pm$ 1.1 & 0.07    & $\pm$ 0.29 & \multicolumn{3}{c}{-} & $< 10^{-7}$&~\cite{BarShalom:2010qr} \\
\end{tabular}
\end{adjustbox}
\caption{\label{Tab:inter:CP_comp} Comparison of measurements of charge asymmetries and constraints on CP asymmetries, with MC simulation (detailed in the text), existing experimental limits and SM predictions. The latter two columns represent upper limits on the absolute values $\lvert A\rvert$.  For $A^{b}_{\mathrm{mix}}$ the last two columns are determined using the prescription from Ref.~\cite{Gedalia:2012sx}, with inputs from the HFAG world average high-energy $f_{d,s}$~\cite{Amhis:2014hma} and either the world average~\cite{Amhis:2014hma} or the SM predictions~\cite{Artuso:2015swg} for $a^{d}_{\mathrm{sl}}$ and $a^{s}_{\mathrm{sl}}$ respectively.}
\end{center}
\end{table}

\FloatBarrier

\section{Conclusions} 
\label{sec:Conclusion}

Same- and opposite-sign charge asymmetries are measured with the ATLAS detector at the LHC in $\ell$+jets~\ttbar{} events using the 2012 data sample corresponding to 20.3~\ifb{} of proton--proton collisions at $\sqrt{s}=8~\tev{}$.
The charge asymmetries are formed from the charge of the lepton from the top-quark decay and from the charge of the soft muon from the semileptonic decay of a $b$-hadron. The lepton from the top-quark decay determines the charge of the produced $b$-quark whilst the charge of the soft lepton determines the $b$-quark charge at decay. The same- or different-top experimental ambiguity is resolved by a kinematic likelihood fitter with a misassignment probability of $21\pm1\%$. Backgrounds are subtracted from the data, which are unfolded to a well-defined fiducial region in which the charge asymmetries are measured. The decay-chain fractions are taken from simulation in the fiducial region and the CP asymmetries are extracted.

This paper presents a measurement of $A^{bc}_{\mathrm{dir}}$, strengthens the existing $2\sigma$ limit on $A^{c\ell}_{\mathrm{dir}}$ and provides an equivalent $2\sigma$ limit on $A^{b\ell}_{\mathrm{dir}}$. All reported results are found to be consistent with the Standard Model. The largest uncertainty on all reported asymmetries is statistical. With the existing 2015-16 Run 2 ATLAS dataset, the statistical uncertainty will already be smaller than the systematic uncertainties.


\section*{Acknowledgements}
We thank CERN for the very successful operation of the LHC, as well as the
support staff from our institutions without whom ATLAS could not be
operated efficiently.

We acknowledge the support of ANPCyT, Argentina; YerPhI, Armenia; ARC, Australia; BMWFW and FWF, Austria; ANAS, Azerbaijan; SSTC, Belarus; CNPq and FAPESP, Brazil; NSERC, NRC and CFI, Canada; CERN; CONICYT, Chile; CAS, MOST and NSFC, China; COLCIENCIAS, Colombia; MSMT CR, MPO CR and VSC CR, Czech Republic; DNRF and DNSRC, Denmark; IN2P3-CNRS, CEA-DSM/IRFU, France; GNSF, Georgia; BMBF, HGF, and MPG, Germany; GSRT, Greece; RGC, Hong Kong SAR, China; ISF, I-CORE and Benoziyo Center, Israel; INFN, Italy; MEXT and JSPS, Japan; CNRST, Morocco; FOM and NWO, Netherlands; RCN, Norway; MNiSW and NCN, Poland; FCT, Portugal; MNE/IFA, Romania; MES of Russia and NRC KI, Russian Federation; JINR; MESTD, Serbia; MSSR, Slovakia; ARRS and MIZ\v{S}, Slovenia; DST/NRF, South Africa; MINECO, Spain; SRC and Wallenberg Foundation, Sweden; SERI, SNSF and Cantons of Bern and Geneva, Switzerland; MOST, Taiwan; TAEK, Turkey; STFC, United Kingdom; DOE and NSF, United States of America. In addition, individual groups and members have received support from BCKDF, the Canada Council, CANARIE, CRC, Compute Canada, FQRNT, and the Ontario Innovation Trust, Canada; EPLANET, ERC, FP7, Horizon 2020 and Marie Sk{\l}odowska-Curie Actions, European Union; Investissements d'Avenir Labex and Idex, ANR, R{\'e}gion Auvergne and Fondation Partager le Savoir, France; DFG and AvH Foundation, Germany; Herakleitos, Thales and Aristeia programmes co-financed by EU-ESF and the Greek NSRF; BSF, GIF and Minerva, Israel; BRF, Norway; Generalitat de Catalunya, Generalitat Valenciana, Spain; the Royal Society and Leverhulme Trust, United Kingdom.

The crucial computing support from all WLCG partners is acknowledged gratefully, in particular from CERN, the ATLAS Tier-1 facilities at TRIUMF (Canada), NDGF (Denmark, Norway, Sweden), CC-IN2P3 (France), KIT/GridKA (Germany), INFN-CNAF (Italy), NL-T1 (Netherlands), PIC (Spain), ASGC (Taiwan), RAL (UK) and BNL (USA), the Tier-2 facilities worldwide and large non-WLCG resource providers. Major contributors of computing resources are listed in Ref.~\cite{ATL-GEN-PUB-2016-002}.

\clearpage
\printbibliography

\newpage 
\begin{flushleft}
{\Large The ATLAS Collaboration}

\bigskip

M.~Aaboud$^\textrm{\scriptsize 137d}$,
G.~Aad$^\textrm{\scriptsize 88}$,
B.~Abbott$^\textrm{\scriptsize 115}$,
J.~Abdallah$^\textrm{\scriptsize 8}$,
O.~Abdinov$^\textrm{\scriptsize 12}$,
B.~Abeloos$^\textrm{\scriptsize 119}$,
O.S.~AbouZeid$^\textrm{\scriptsize 139}$,
N.L.~Abraham$^\textrm{\scriptsize 151}$,
H.~Abramowicz$^\textrm{\scriptsize 155}$,
H.~Abreu$^\textrm{\scriptsize 154}$
% ...
\bigskip
\\
% ...
$^{8}$ Department of Physics, The University of Texas at Arlington, Arlington TX, United States of America\\
% ...
$^{12}$ Institute of Physics, Azerbaijan Academy of Sciences, Baku, Azerbaijan\\
% ...
$^{88}$ CPPM, Aix-Marseille Universit{\'e} and CNRS/IN2P3, Marseille, France\\
% ...
$^{115}$ Homer L. Dodge Department of Physics and Astronomy, University of Oklahoma, Norman OK, United States of America\\
% ...
$^{119}$ LAL, Univ. Paris-Sud, CNRS/IN2P3, Universit{\'e} Paris-Saclay, Orsay, France\\
% ...
$^{137}$ $^{(a)}$ Facult{\'e} des Sciences Ain Chock, R{\'e}seau Universitaire de Physique des Hautes Energies - Universit{\'e} Hassan II, Casablanca; $^{(b)}$ Centre National de l'Energie des Sciences Techniques Nucleaires, Rabat; $^{(c)}$ Facult{\'e} des Sciences Semlalia, Universit{\'e} Cadi Ayyad, LPHEA-Marrakech; $^{(d)}$ Facult{\'e} des Sciences, Universit{\'e} Mohamed Premier and LPTPM, Oujda; $^{(e)}$ Facult{\'e} des sciences, Universit{\'e} Mohammed V, Rabat, Morocco\\
% ...
$^{139}$ Santa Cruz Institute for Particle Physics, University of California Santa Cruz, Santa Cruz CA, United States of America\\
% ...
$^{151}$ Department of Physics and Astronomy, University of Sussex, Brighton, United Kingdom\\
% ...
$^{154}$ Department of Physics, Technion: Israel Institute of Technology, Haifa, Israel\\
$^{155}$ Raymond and Beverly Sackler School of Physics and Astronomy, Tel Aviv University, Tel Aviv, Israel\\
% ...
$^{a}$ Also at Department of Physics, King's College London, London, United Kingdom\\
$^{b}$ Also at Institute of Physics, Azerbaijan Academy of Sciences, Baku, Azerbaijan\\
$^{c}$ Also at Novosibirsk State University, Novosibirsk, Russia\\
$^{d}$ Also at TRIUMF, Vancouver BC, Canada\\
$^{e}$ Also at Department of Physics {\&} Astronomy, University of Louisville, Louisville, KY, United States of America\\
% ...
\end{flushleft}

\end{document}